\documentclass[useAMS,usegraphicx,usenatbib]{mn2e}
\usepackage{threeparttable}
\usepackage{times}
\usepackage{epsfig}
\usepackage{textcomp}
\usepackage{amssymb}
\usepackage{subfigure}
\usepackage{multirow}
\usepackage{enumerate}

\title[SWIRE 70 $\mu$m selected galaxies]{Spectroscopic follow-up of 70 $\mu$m sources in \textit{Spitzer} Wide-area Infrared Extragalactic Legacy Survey}

\author[H.Patel et~al.]{H. Patel$^1$$\thanks{harsit.patel08@imperial.ac.uk}$, D. L. Clements$^1$, M. Rowan-Robinson$^1$, M. Vaccari$^2$\\
$^1$Astrophysics Group, Imperial College London, Blackett Laboratory, Prince Consort Road, London SW7~2AW\\
$^2$Dipartimento di Astronomia, Universit\`{a} di Padova, vicolo Osservatorio, 3, 35122 Padova, Italy}

\begin{document}


\date{Accepted. Received}

\pagerange{\pageref{firstpage}--\pageref{lastpage}} \pubyear{2010}

\maketitle

\label{firstpage}

\begin{abstract}
We present spectroscopic follow-up observations of 70 $\mu$m selected galaxies from the SWIRE XMMLSS and Lockman Hole fields. We have measured spectroscopic redshifts for 293 new sources down to a 70 $\mu$m flux limit of 9mJy and  $r<$ 22 mag. The redshift distribution peaks at $z\sim$ 0.3 and has a high redshift tail out to $z=3.5$. We perform emission line diagnostics for 91 sources where  [OIII], H$\beta$, [NII], H$\alpha$ and [SII] emission lines are available to determine their power source. We find in our sample 13 QSOs, 1 Seyfert II galaxy, 33 star forming galaxies, 30 composite galaxies, 5 LINERs and 21 ambiguous galaxies. We fit single temperature dust spectral energy distributions (SEDs) to 81 70 $\mu$m sources with 160 $\mu$m photometry to estimate dust temperatures and masses.  Assuming the dust emissivity factor ($\beta$) as 1.5, we determine dust temperatures in the range $\sim$ 20-60K and dust masses with a range of 10$^6$-10$^9$ M$_\odot$. Plotting these objects in the luminosity-temperature diagram suggests that these objects have lower dust temperatures than local IR luminous galaxies. The \textit{Herschel} Space Observatory will be crucial in understanding the nature of these sources and to accurately determining the shape of the Rayleigh-Jeans tail of the dust SED. We then model SEDs from optical to far-IR for each source using a set of galaxy and quasar templates in the optical and near-IR (NIR) and with a set of dust emission templates (cirrus, M82 starburst, Arp 220 starburst and AGN dust torus) in the mid-IR (MIR) to far-IR (FIR). The number of objects fit with each dust template are: 57 Arp 220, 127 M82, 9 cirrus, 1 AGN dust torus, 70 M82 and cirrus, 26 M82 and AGN dust torus and 3 Arp 220 and AGN dust torus. We determine the total IR luminosity (L$_\mathrm{IR}$) in range 10$^8$-10$^{15}$ L$_\odot$ by integrating the SED models from 8 to 1000 $\mu$m.
\end{abstract}


\section{INTRODUCTION}
The discovery of the Cosmic Infrared Background (CIRB; \citealt{puget96, fixsen98} has shown that at least half of the energy generated by star-formation and Active Galactic Nuclei (AGN) in the Universe has been absorbed by dust and re-radiated in the IR \citep{gispert00,hauser01}. A significant fraction of the background has been resolved into a population of IR sources, luminous (LIRG:  L$_\mathrm{IR}$  = 10$^{11}$-10$^{12}$  L$_\odot$),  ultraluminous (ULIRG: L$_\mathrm{IR}$  = 10$^{12}$-10$^{13}$  L$_\odot$) and hyperluminous (HLIRG:  L$_\mathrm{IR}  >$ 10$^{13}$  L$_\odot$) infrared (IR) galaxies whose bolometric energy output is dominated in the IR ($\lambda$ = 8-1000 $\mu$m) by reprocessed dust emission (see \citealt{sanders96, chary01, franceschini01}). They were first catalogued in the local Universe ($z\leq$ 0.3) by the \textit{InfraRed Astronomical Satellite}  (IRAS) and shown to only contribute $\sim$ 30\% to the CIRB which suggests that the majority originates from reprocessed dust emission by high redshift galaxies.

The \textit{Spitzer Space Telescope} \citep{werner04} has greatly increased our understanding of the IR Universe at high redshifts. We now know that there is strong evolution in the population of IR galaxies with LIRGs dominating the cosmic star formation rate out to $z\sim1$ and ULIRGs out to $z\sim2$ \citep{lefloch05, perezgonzalez05, caputi07}. This evolution is also seen in the more luminous submillimeter galaxies (SMGs) at redshifts $>$ 2 \citep{smail97,scott02, clements08, dye08}. These studies have highlighted that a large fraction of star formation and gravitational accretion is heavily obscured by dust and therefore studying the properties and the nature of these galaxies is crucial to understanding galaxy evolution.

\indent Deep surveys conducted by \textit{Spitzer}, particularly in the mid-IR (MIR) at 24 $\mu$m where the MIPS instrument is most sensitive have allowed us to understand the nature of dust obscured galaxies. Using spectroscopic diagnostics and/or modelling the spectral energy distributions (SEDs) combining optical and IR photometry has shown that star-formation is the dominant process with the AGN component becoming significant in the more luminous sources \citep{genzel00, farrah03, soifer08}. However the 24 $\mu$m surveys are limited as the peak of the SED for most IR luminous galaxies is in the range of $\sim 40 - 200$ $\mu$m. At high redshifts the 24 $\mu$m channel samples shorter wavelength regions contaminated by polycyclic aromatic hydrocarbon (PAH) emission and silicate absorption features making it difficult to obtain reliable estimates of the bolometric IR luminosities.

\indent \textit{Spitzer} also observed at the longer 70 and 160 $\mu$m wavelengths and several studies have been carried out in order to characterise the 70 $\mu$m population. Analysis of the 160 $\mu$m population is severely hindered because of the lower resolution and sensitivity of the MIPS instrument at this wavelength. By modelling the IR SEDs of 70 $\mu$m selected galaxies in the $0.1 < z < 2$ redshift range, \cite{symeonidis07, symeonidis08} found that these objects require a cold (dust temperature $<$ 20K) emission component in the FIR. In the follow-up study, \cite{symeonidis09} (henceforth S09) suggest that the MIPS 70 $\mu$m population may be the missing link between the cold $z>1$ submillimeter (submm) population detected by SCUBA and the local IR luminous galaxies. \cite{kartaltepe10} extend their study to higher redshifts, $z\sim 3.5$ for 1503 reliable and unconfused 70 $\mu$m selected sources in the Cosmic Evolution Survey (COSMOS). Their analysis shows that the SED shapes are similar to local objects but also find evidence for a cooler component than is observed locally. Using ancillary radio and X-ray data, they find that the fraction of AGNs increases with L$_\mathrm{IR}$, and nearly 51\% of ULIRGs and all HLIRGs likely host a powerful AGN. \cite{symeonidis10} also arrive at the same conclusion although they find 23\% of ULIRGs contain an AGN. These figures are consistent with observations of local ULRIGs (see \citealt{veilleux95, sanders96, kim98}) which show that the AGN fraction increases from $\sim$ 4 \% at L$_\mathrm{IR}$ = 10$^{10}$ L$_\odot$ to $>$ 50\% at L$_\mathrm{IR} > $  10$^{12}$ L$_\odot$. 

\indent In this paper, we examine the optical and IR properties from a spectroscopic follow-up of 293 70 $\mu$m selected sources from the XMM-LSS and Lockman Hole (LH) region of the \textit{Spitzer} Wide-area InfraRed Extragalactic survey (SWIRE) \citep{lonsdale03}. The present paper is the first in a series which aims to study the evolution of the 70 $\mu$m luminosity function. This paper is organised as follows: In Sect.\ref{sect:data} we present the sample selection criteria and in Sect.\ref{sect:obsdatareduc} we describe the observations and data reduction method. We present the main results in Sect.\ref{sect:results} and discuss the implications in Sect.\ref{sect:summary}. Throughout this paper, we assume a $\Lambda$CDM cosmology with $H_\mathrm{0}$ = 70kms$^{-1}$Mpc$^{-1}$, $\Omega_{\Lambda}$ = 0.7, and $\Omega_\mathrm{M}$ = 0.3. All magnitudes are in the AB system unless otherwise stated.

\section{THE DATA}
\label{sect:data}
In this paper we have used data from the latest release of the SWIRE photometric redshift catalogue of \cite{mrr08}. The catalogue contains photometric redshifts for over 1 million IR sources, estimated by combining the optical and IRAC 3.6 and 4.5 $\mu$m photometry to fit the observed SEDs with a combination of galaxy and AGN templates \citep{babbedge04, mrr05}. Specifically, we make use of multiwavelength data from XMM-LSS and LH-ROSAT regions of the SWIRE survey.

\subsection{Infrared data}
\label{sect:irdata}
\indent The SWIRE survey \citep{lonsdale03, lonsdale04} is one of the largest \textit{Spitzer} legacy programmes covering 49 deg$^2$ in six different fields with both the IRAC and MIPS instruments. Typical 5$\sigma$ sensitivities are 3.7, 5.3, 48 and 37.7 $\mu$Jy in the IRAC 3.6, 4.5, 5.8 and 8 $\mu$m bands. For MIPS the 5$\sigma$ limits are  230$\mu$Jy, 20 mJy, and 120 mJy at 24, 70 and 160 $\mu$m \citep{surace05}. The IRAC and MIPS data were processed by the \textit{Spitzer} Science Centre (SSC) and sources extracted using SExtractor \citep{bertin1996117393}. Full details of the SWIRE data release can be found in \cite{surace05}.

\indent The final data product consists of a bandmerged IRAC and MIPS 24 $\mu$m catalog and single-band catalogs at 24, 70 and 160 $\mu$m. The IRAC and MIPS 24 $\mu$m catalog consists of sources detected with a signal-to-noise ratio (S/N) $>$ 5 in one or more IRAC bands and their 24 $\mu$m associations with a S/N $>$ 3. The MIPS 70 and 160 $\mu$m sources single-band catalogs were matched to the IRAC and MIPS 24 $\mu$m bandmerged catalogs to produce a SWIRE IRAC and MIPS 7-band catalog. Thus, almost all of the 70 and 160 $\mu$m sources are also detected at 24 $\mu$m.

\subsection{Optical data}
\label{sect:opticaldata}
Optical photometry is available for $>$ 70\% of the SWIRE area in at least three of the \textit{U}, \textit{g'}, \textit{r'}, \textit{i'} and \textit{Z} photometric bands. Spitzer-Optical cross-identifications (XID) was carried out  between the optical and the IRAC-24 $\mu$m catalogs using a search radius of 1.5$''$ \citep{mrr05, surace05}. The cross-identification process ensured that each SWIRE source only had one optical match. Completeness and reliability of the XID was investigated by \cite{surace05}, which showed that essentially the \textit{Spitzer}-optical XIDs are essentially 100\% complete. The requirement that appears to eliminate spurious sources effectively and give a high-reliability catalog is that the source be detected at both 3.6 and 4.5 $\mu$m at S/N $\geq$ 5 \citep{mrr05, surace05}.

\subsubsection{XMM-LSS}
\label{sect:xmmlss}
The XMM-LSS centred on $\alpha$ = 02$^\mathrm{h}$21$^\mathrm{m}$20$^\mathrm{s}$, $\delta$ = -04$^\mathrm{d}$30$^\mathrm{m}$00$^\mathrm{s}$ covers 9.1 deg$^2$. Optical photometry is available for 6.97 deg$^2$ of XMM-LSS, which was observed as part of the Canda-France-Hawaii Telescope Legacy Survey (CFHTLS) in the \textit{u*, g$^\prime$, r$^\prime$, i$^\prime$}, and \textit{z`} bands to magnitude (Vega, 5$\sigma$ for a point like object) limits of: 24.9, 26.4, 25.5, 24.9 and 23.4 respectively\footnote{See http://www.cfht.hawaii.edu/Science/CFHTLS/}. The photometry was obtained from \cite{pierre07}. 

\subsubsection{LH-ROSAT}
\label{sect:lhrosat}
The LH-ROSAT field is smaller part of the larger LH region centred on $\alpha$ = 10$^\mathrm{h}$52$^\mathrm{m}$43$^\mathrm{s}$, $\delta$ = +57$^\mathrm{d}$28$^\mathrm{m}$48$^\mathrm{s}$ and covers 0.25 deg$^2$. Optical photometry in the \textit{U, g`, r`} and \textit{i`} bands were obtained using the MOSAIC camera on the 4m-Mayall Telescope at Kitt Peak National Observatory (KPNO). The 5$\sigma$ limiting magnitudes (Vega) are 24.1, 25.1, 24.4, and 23.7 in the four bands for point-like sources \citep{berta07}.

\subsection{Sample selection}
\label{sect:targetselect}
We selected 70 $\mu$m sources with \textit{r} $<$ 22 mag, S$_{70} > $ 9mJy and SNR $>$ 5 at 70 $\mu$m for the spectroscopic follow-up. We found 1553 70 $\mu$m sources in XMM-LSS satisfying the selection criteria, of which 22 had a measured spectroscopic redshift. In LH-ROSAT 58 sources were found, 13 of which also had measured spectroscopic redshift. Thus, our final sample contained 1553 and 45 70 $\mu$m sources in XMM-LSS and LH-ROSAT respectively. 

\subsection{Fibre configuration}
\label{sect:maskfilter}
Fibre configuration was carried out using the \textsf{AF2\_configure}\footnote{See http://www.ing.iac.es/Astronomy/instruments/af2/configuration.html} programme. In total there are 150 science fibres and 10 fiducial fibres which are used to field acquisition and guiding. On average fibres were placed on $\sim$ 70 primary targets (70 $\mu$m sources) and at least 4 standard stars per mask. In addition 10-20 fibres were used for simultaneous sky observations which are referred to as sky fibres. Spare fibres were placed on 24 $\mu$m sources in particular high redshift ($z_\mathrm{ph}>3$) candidates in the XMM-LSS and LH-ROSAT fields, the results of which will be presented in a future paper (Hyde et al. in prep). Note that due to fibre collision, it was not possible to utilise all the science fibres.

\begin{figure*}
\begin{center}
\vspace{0.2cm}
\hspace{-0.5cm}
\subfigure{\label{fig:zdist}\includegraphics[height=4.5cm, width=7cm]{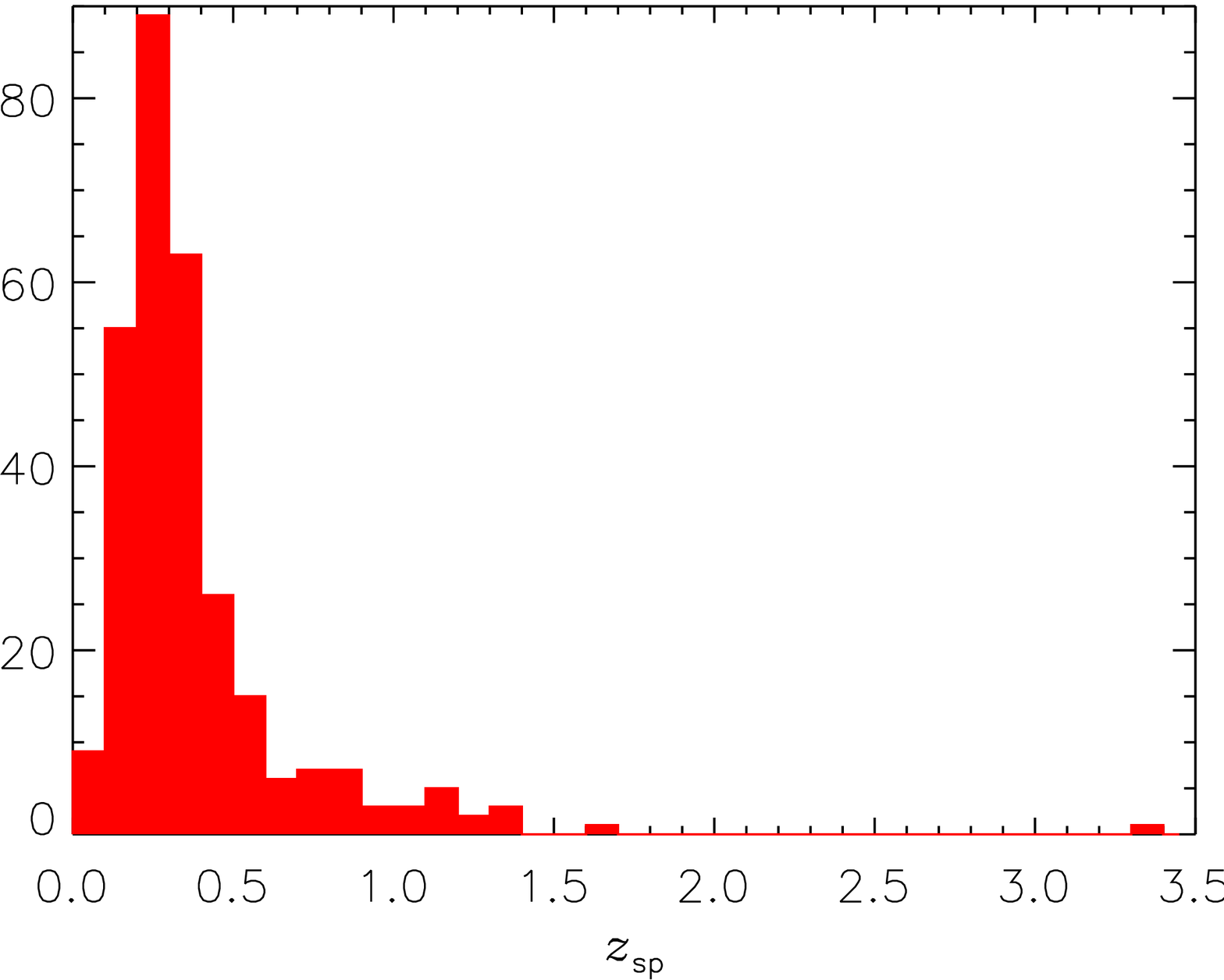}}
\hspace{0.5cm}
\subfigure{\label{fig:rdist}\includegraphics[height=4.5cm, width=7cm]{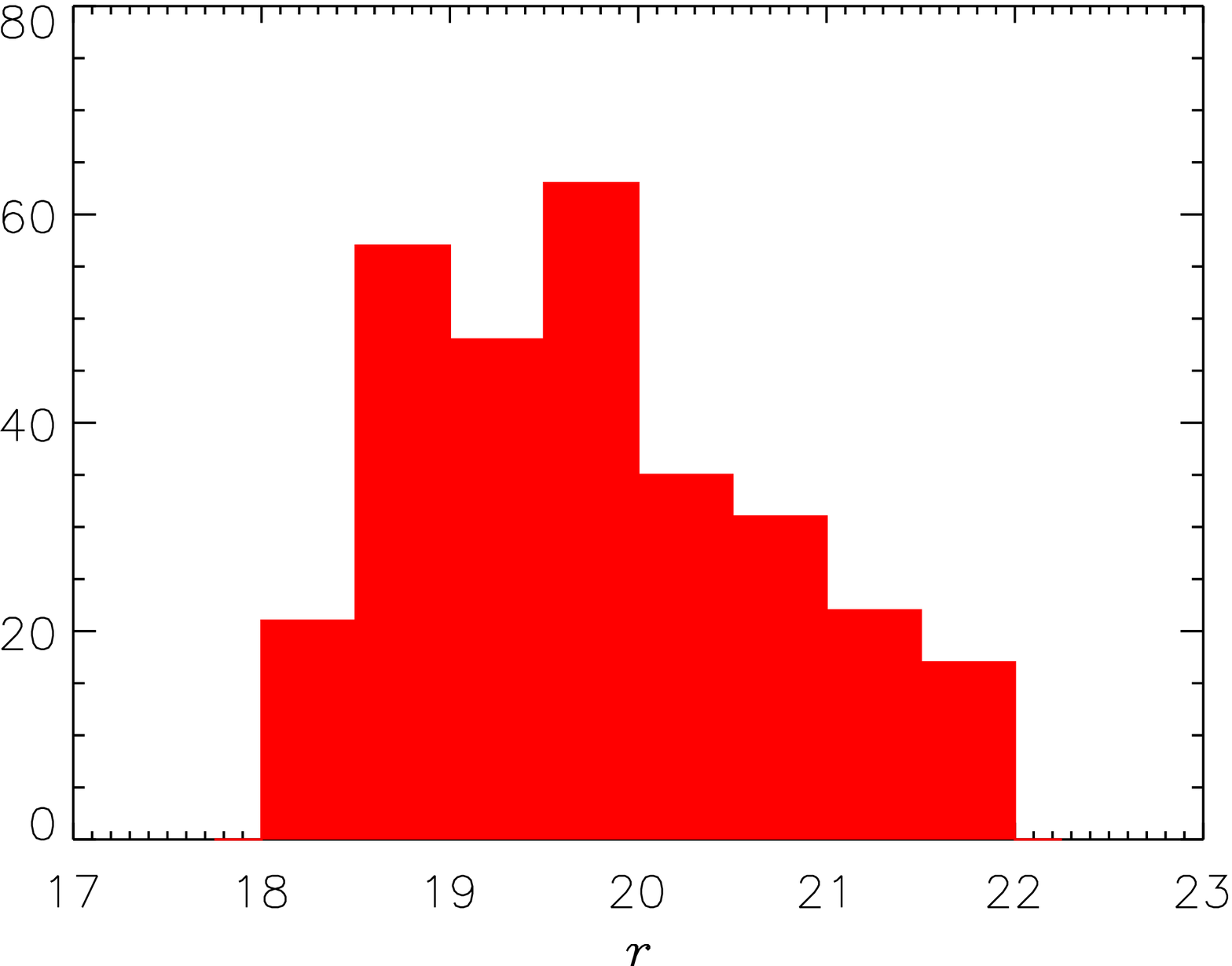}}
\end{center}
\caption{\textbf{Left}: Spectroscopic redshift distribution. \textbf{Right}: \textit{r} magnitude distribution.}
\label{fig:zrdist}
\end{figure*}

\begin{figure*}
\begin{center}
\vspace{1.5cm}
\hspace{1.3cm}
\includegraphics[width=16.2cm, height = 14.9cm]{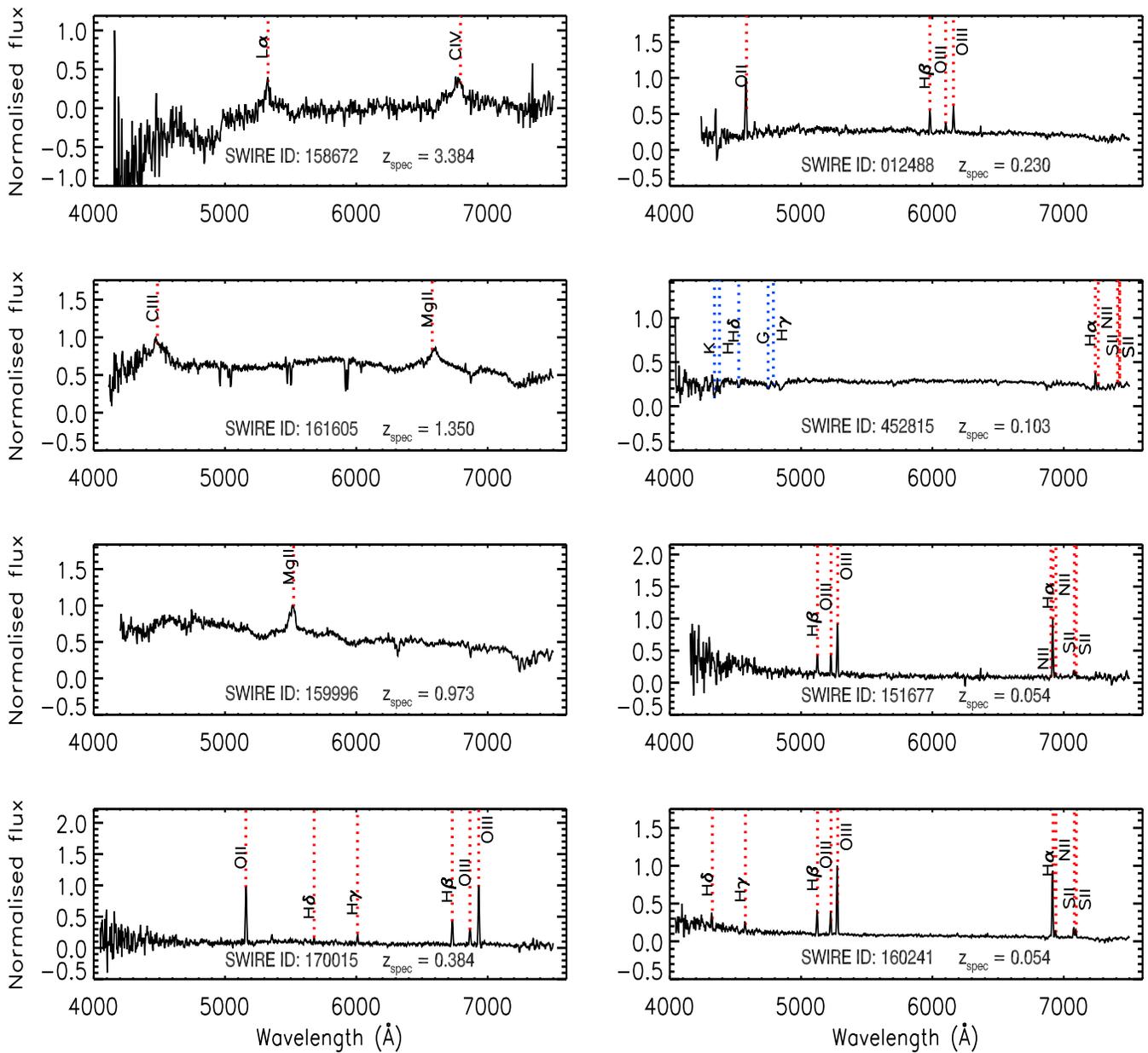}
\end{center}
\caption{Selected example optical spectra of 70 $\mu$m sources observed at WHT using AF2/WYFFOS.}
\label{fig:wht70muspectra}
\end{figure*}

\section{Observations and Data Reduction}
\label{sect:obsdatareduc}
The targets were observed using the Wide Field Fibre Optical Spectrograph (WYFFOS) and the robot positioner Autofib2 at the Observatorio del Roque de los Muchachos, using the 4.2m William Herschel Telescope (WHT). We carried out the observations over 2 runs totalling 6 nights between November 2008-2009 using the small fibre bundles (fibre diameter 1.6 arcsec) and the R316R grating covering the spectral range $\sim$ 4000-9500$\mathrm{\AA}$ and a spectral resolution of 8.1\AA. Each fibre configuration was observed for 3 hours giving a S/N of $\sim$ 5$\sigma$ per resolution element for sources with $r=22$ mag. Bias, and dark frames were obtained at the start and end of the night. Dome flats and arc lamps (helium and neon) were obtained at the start and end of each observed field. Standard stars were also observed at the start and end of the night for flux calibration.
  
\indent The spectra were reduced and calibrated using standard IRAF routines and the WYFFOS/AF2 reduction tool made available online. The tool is based on the \textsf{dofibre} package which was used to remove cosmic-rays, subtract the bias and dark frames, flat-field, and wavelength calibrate, producing 1D spectra. The sky fibres (referred to as sky spectra) used for the sky correction procedure were median combined to create a master sky spectrum. A correction factor was then manually determined by comparing the brightest sky lines (the OI 5570$\mathrm{\AA}$ or the NaD 5890$\mathrm{\AA}$ atmospheric emission lines) in the target and sky spectra. The sky subtracted target spectra (referred to as object spectra) were determined by subtracting the master sky spectrum corrected for the factor from the target spectra. All object spectra were examined visually to validate the sky subtraction accuracy. Standard star frames were reduced using the same method and the object spectra calibrated using the IRAF \textsf{standard} and \textsf{calibrate} packages and atmospheric extinction corrected for the La Palma site. Redshifts were measured using the IRAF package \textsf{rvsao} and by visual inspection of the sky subtracted and calibrated 1D spectra by identifying emission and absorption lines.  A summary of the observation log and spectroscopic redshift success is given in Table \ref{tab:obslog}.

\begin{table*}
\begin{center}
\begin{tabular}{cccccccc}
\hline
Fibre configuration & Observation Date & $\alpha$ (J2000) & $\delta$ (J2000)  & \multicolumn{2} {c} {Sources observed}  & \multicolumn{2} {c}{Redshift measured }\\
& & & & 24 $\mu$m & 70 $\mu$m & 24 $\mu$m & 70 $\mu$m\\
\hline
XMM-LSS\_1 & 23/11/08 & 02$^\mathrm{h}$18$^\mathrm{m}$44$^\mathrm{s}$ & -05$^\mathrm{d}$56$^\mathrm{m}$24$^\mathrm{s}$ & 12 & 80 & 5 & 23 \\
XMM-LSS\_2 & 23/11/08-24/11/08 & 02$^\mathrm{h}$21$^\mathrm{m}$26$^\mathrm{s}$ & -05$^\mathrm{d}$37$^\mathrm{m}$48$^\mathrm{s}$ & 9 &69 & 4 & 25\\
XMM-LSS\_3 & 24//11/08 & 02$^\mathrm{h}$24$^\mathrm{m}$26$^\mathrm{s}$ & -05$^\mathrm{d}$24$^\mathrm{m}$00$^\mathrm{s}$ & 11 & 74 & 3 & 46\\
XMM-LSS\_4 & 24/11/08-18/11/09 & 02$^\mathrm{h}$27$^\mathrm{m}$09$^\mathrm{s}$ & -05$^\mathrm{d}$15$^\mathrm{m}$50$^\mathrm{s}$ & 9 & 77 & 4 & 52\\
XMM-LSS\_5 & 18/11/09 & 02$^\mathrm{h}$22$^\mathrm{m}$31$^\mathrm{s}$ & -04$^\mathrm{d}$50$^\mathrm{m}$24$^\mathrm{s}$ & 13 & 76 & 6 & 46\\
XMM-LSS\_6 & 18/11/09-19/11/09 & 02$^\mathrm{h}$18$^\mathrm{m}$48$^\mathrm{s}$ & -05$^\mathrm{d}$42$^\mathrm{m}$00$^\mathrm{s}$  & 14 & 72 & 7 & 45\\
XMM-LSS\_7 & 19/11/09 & 02$^\mathrm{h}$22$^\mathrm{m}$34$^\mathrm{s}$ & -04$^\mathrm{d}$08$^\mathrm{m}$56$^\mathrm{s}$ & 6 & 74 & 4 & 44\\
LH-ROSAT\_1& 18/11/09-19/11/09 & 10$^\mathrm{h}$52$^\mathrm{m}$43$^\mathrm{s}$ & +57$^\mathrm{d}$25$^\mathrm{m}$12$^\mathrm{s}$ & 3 & 24 & 2 & 13\\
\hline
\end{tabular}
\end{center}
\label{tab:obslog}
\caption{Summary of the observation log. The RA and DEC are the centre of each mask. Column 5 lists the number of 70 $\mu$m sources targeted and column 6 is the number of successful spectroscopic redshift measured in each pointing.}
\end{table*}

\begin{figure*}
\begin{center}
\vspace{1.5cm}
\hspace{1.3cm}
\includegraphics[width=16.cm, height = 14.9cm]{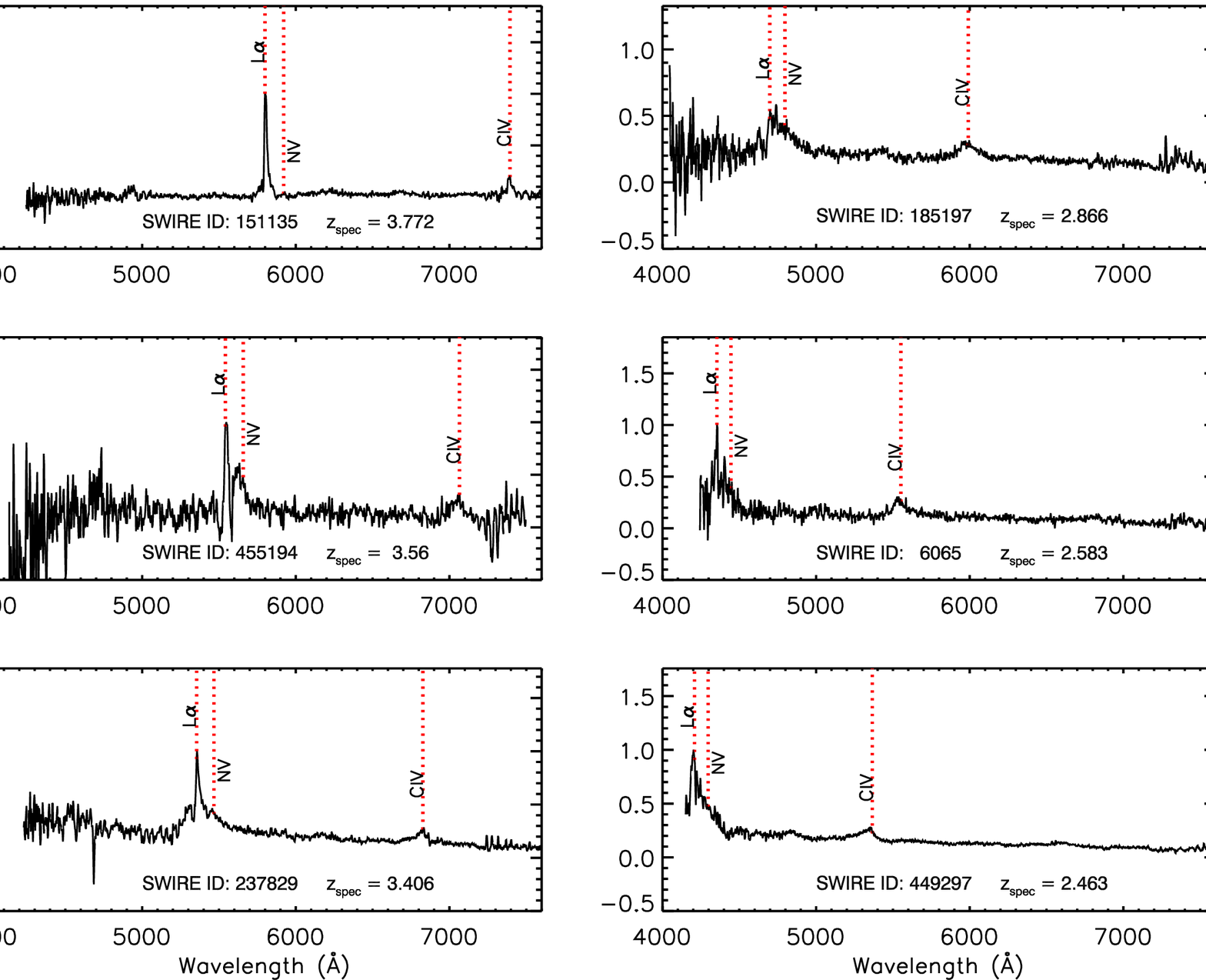}
\end{center}
\caption{Selected example optical spectra of 24$\mu$m sources observed at WHT using AF2/WYFFOS.}
\label{fig:wht24muspectra}
\end{figure*}

\begin{figure*}
\begin{center}
\vspace{0.5cm}
\hspace{0.1cm}
\subfigure{\label{fig:zphvzsp24um}\includegraphics[width=7.6cm, height = 6.2cm]{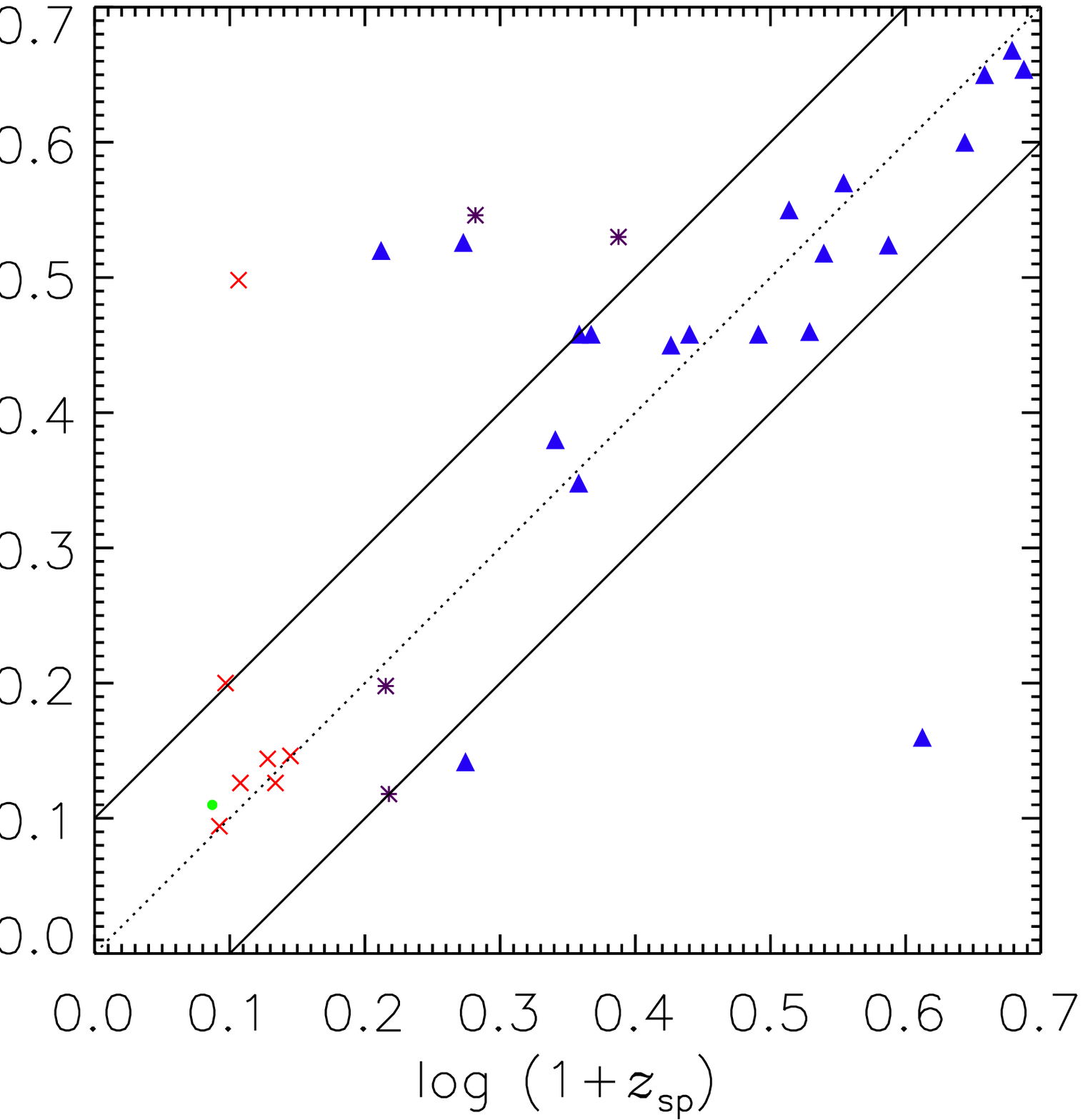}}
\hspace{0.7cm}
\subfigure{\label{fig:zphvzsp}\includegraphics[width=7.6cm, height = 6.2cm]{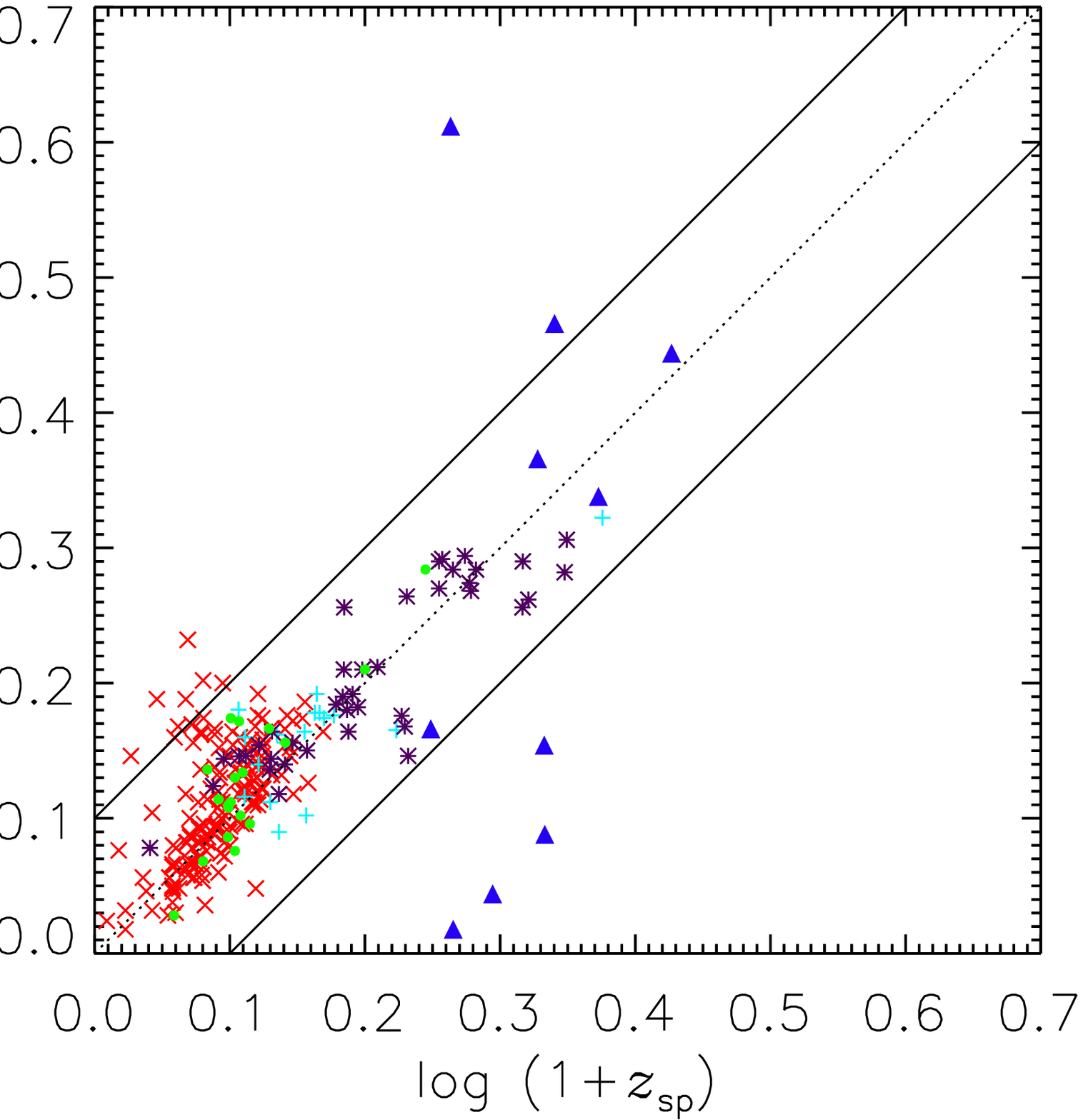}}
\end{center}
\caption{Comparison of spectroscopic against photometric redshift for objects with $\chi^2_\mathrm{red}<10$ from the catalogue of \citet{mrr08} for 24 (left panel) and 70 $\mu$m (right panel) sources. The two solid lines are the 10\% error boundary in $\log(1+z)$. The data are plotted according to the redshift quality flag, (zflag; see text in Sect.\ref{sect:results}), where red crosses are objects with zflag = 0, magenta pluses are objects with zflag = 1, purple crosses are objects with zflag = 2 and green circles are objects with zflag = 3. Blue triangles are the QSOs identified from spectra.}
\label{fig:zcomp}
\end{figure*}

\vspace{-2.5mm}

\section{Results}
\label{sect:results}
We observed a total of 477 70 $\mu$m sources in the SWIRE XMM-LSS and LH-ROSAT fields and measured spectroscopic redshifts for 293 sources (280 in XMM-LSS and 13 in LH-ROSAT), giving a success rate of $\sim$ 61\%. The first observation run was severely affected by high humidity which resulted in poor quality spectra for the first two fields which explains the low redshift success rate ($\sim$ 36\%). For the second observing run, we have a measured redshift success rate of $\sim$ 71\%. Redshifts for some objects could not be measured due to sky subtraction residuals resulting from poor OH emission line subtraction in the wavelength region $\sim$ 7000 - 9500$\mathrm{\AA}$. A table of our spectroscopic redshift catalogue is provided as online material. 

\indent For each measured redshift, we have assigned a redshift quality flag. The flag has values between 0-3, with classification as follows: zflag = 0, highly reliable redshift with three or more spectral features; zflag = 1: redshift with at least two emission lines; zflag = 2, redshift from one emission line and consistent with the photo-z estimate; zflag = 3, redshift from absorption features only and consistent with photo-z estimate. Our sample contains 194 objects with zflag = 0, 21 with zflag = 1, 59 with zflag = 2 and 19 with zflag = 3. The redshift distribution and the magnitude distribution of the sources in our spectroscopic sample are shown in Fig.\ref{fig:zrdist}. The redshift distribution peaks at $z\sim0.3$ with a high-redshift tail out to $z=3.389$. We show in Fig.\ref{fig:wht70muspectra} and Fig.\ref{fig:wht24muspectra} a selection of optical spectra for a selection of 70 and 24 $\mu$m sources.

\subsection{Photometric redshift comparison}
\label{sect:photozcomp}

We have compared our spectroscopic redshifts with the photometric redshift estimates using the ImpZ code \citep{babbedge04} in the photometric redshift catalogue of \cite{mrr08}. The reliability of the photometric redshifts are measured via the fractional error $\Delta{z}/(1+z)$ for each source, examining the mean error $\bar{\Delta}z/(1+z)$, the total rms scatter $\sigma_z$ and the rate of `catestrophic' outliers, $\eta$, defined as the fraction of the full sample that has $|\Delta{z/(1+z)}| > $ 0.1. 

\indent We show the comparison of spectroscopic redshift ($\log{(1+z_\mathrm{sp})}$) against photometric redshift ($\log{(1+z_\mathrm{ph})}$) in Fig.\ref{fig:zcomp} for objects which have a reliable photometric redshift, that is objects with reduced $\chi^2$ ($\chi^2_\mathrm{red}$) $<$ 10 in the photometric redshift catalogue of \cite{mrr08}. The symbols represent the redshift quality flag while the QSOs (blue triangles) are separated as the photometric redshift solutions are expected to be less accurate for AGNs than galaxies. For the 70 $\mu$m sample (right panel Fig.\ref{fig:zcomp}), the total rms scatter, $\sigma_{z}$ was 0.14, the mean error was 0.035, and the rate of catastrophic outliers $\eta$ was 5.7\%. We find in our 70 $\mu$m sample 14 catastrophic outliers, 6 of which are QSOs and the rest have a reliable redshift with zflag = 0. The 24 $\mu$m sample is characterised by $\sigma_{z}$ = 0.13, the mean error = 0.42 and $\eta$ = 25.8\%. For QSOs as stated, the photometric codes are expected to perform less reliably possibly because of photometric  variability. For the other outliers the causes may be poor optical photometry or photometric redshift aliasing, but overall the low catastrophic outlier rate is encouraging for use of photometric redshifts in future analysis.

\subsection{Emission line diagnostic}
\label{sect:bpt}
We have classified narrow emission line galaxies and determined the dominant power sources using emission line diagnostics or BPT \citep{baldwin81} diagrams. The diagrams are based on optical emission line ratios such as [OIII]/H$\beta$, [NII]/H$\alpha$, [SII]/H$\alpha$ and [OII]/H$\alpha$.  We have used a combination of [OIII]/H$\beta$, [NII]/H$\alpha$ and  [SII]/H$\alpha$ emission line ratios as these ratios are insensitive to reddening because they are close in wavelength. \cite{kewley01} derive a new theoretical `maximum starburst line' on the BPT diagram to separate AGNs and star forming galaxies, with 
galaxies lying above the line likely to be AGN dominated.  \cite{kauffmann03} modified the \cite{kewley01} scheme by including another line which includes objects whose spectra contain both AGN and star formation. Here we use the classification scheme presented in \cite{kewley06} to classify objects in our spectroscopic sample.

\begin{figure*}
\begin{center}
\vspace{0.5cm}
\hspace{0.3cm}
\includegraphics[width=7.5cm, height = 6.5cm]{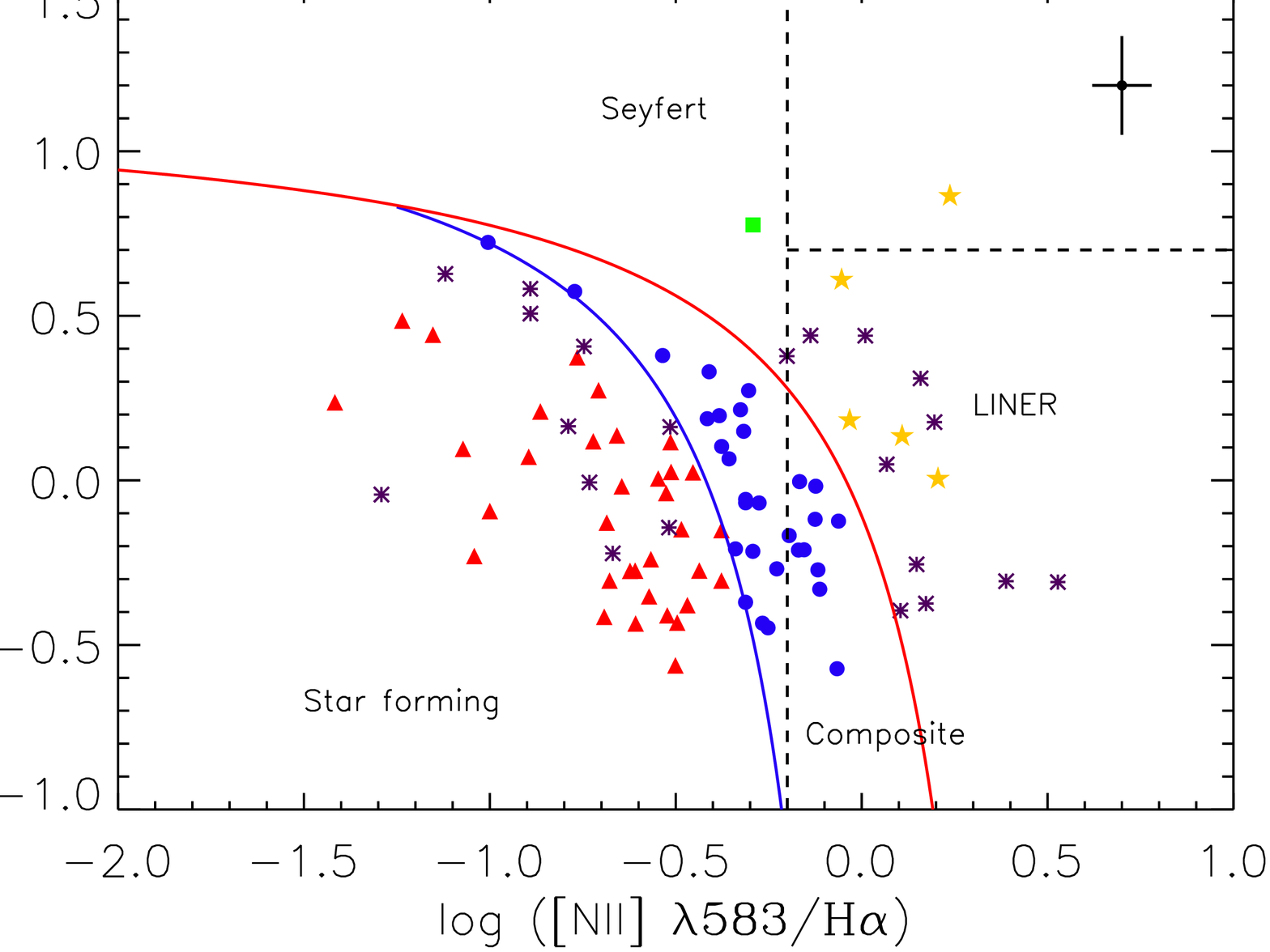}
\hspace{0.7cm}
\includegraphics[height=6.5cm, width=7.5cm]{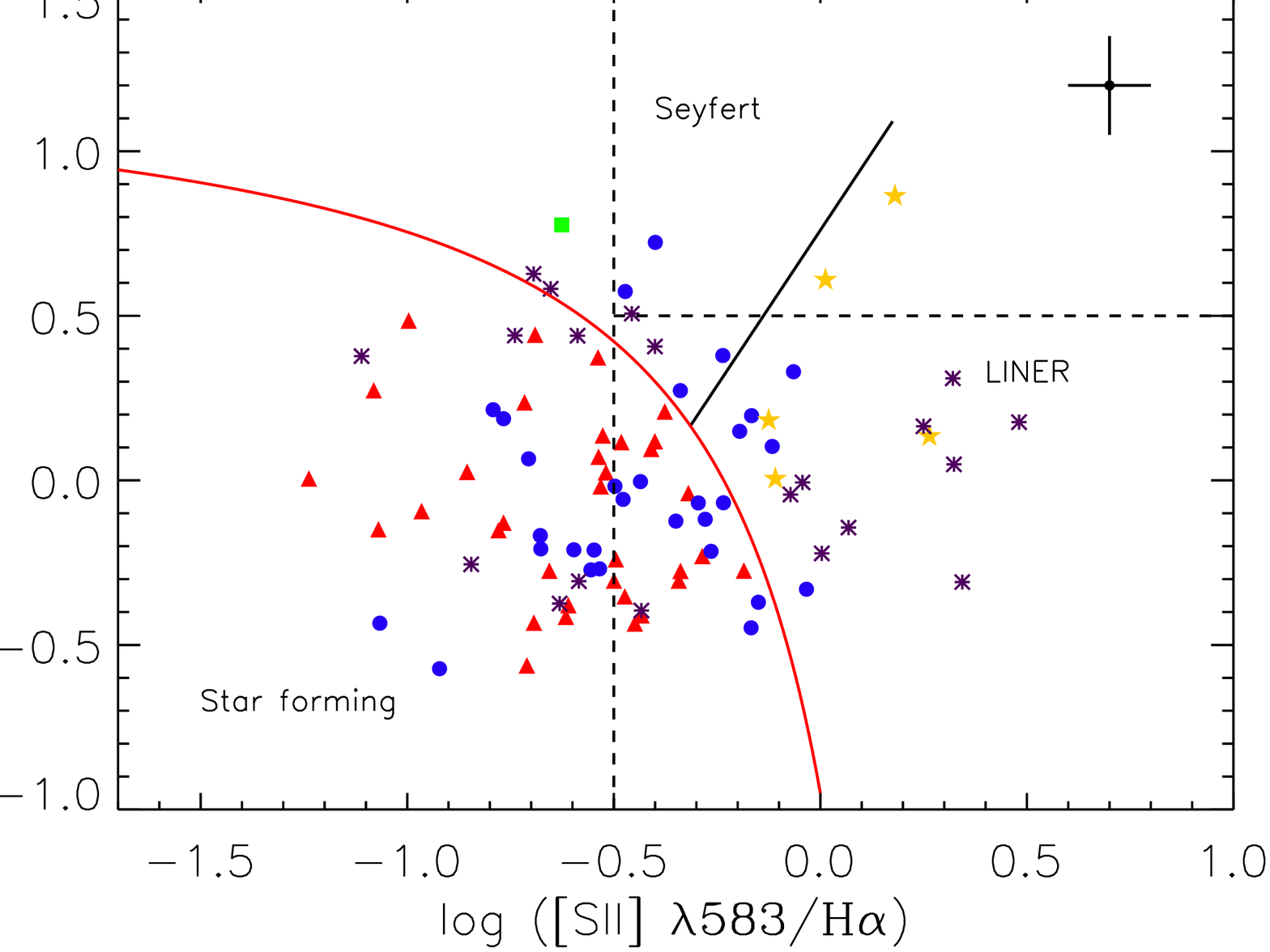}
\end{center}
\caption{\textbf{Left}: [NII]/H$\alpha$ versus [OIII]/H$\beta$ diagnostic diagram. The blue line is the \citet{kauffmann03} pure star formation line and the blue line is the \citet{kewley01} extreme starburst line. \textbf{Right}: [SII]/H$\alpha$ versus [OIII]/H$\beta$ diagnostic diagram. Red line is the extreme starburst of \citet{kewley01} is used to separate the star forming Seyfert galaxies. The black solid line is the Seyfert/LINER separation from \citet{kewley06}. In both the panels the red triangles are pure star forming galaxies, blue circles are composite galaxies, green squares are Seyferts, orange stars are LINERs and purple asterisks are objects classified as ambiguous. The dashed lines represent the \citet{ho97} classification scheme. A typical error bar is shown on the top right hand corner.}
\label{fig:bptdiag}
\end{figure*}

\begin{figure}
\begin{center}
\vspace{0.5cm}
\hspace{0.5cm}
\includegraphics[width=7.2cm, height=5.5cm]{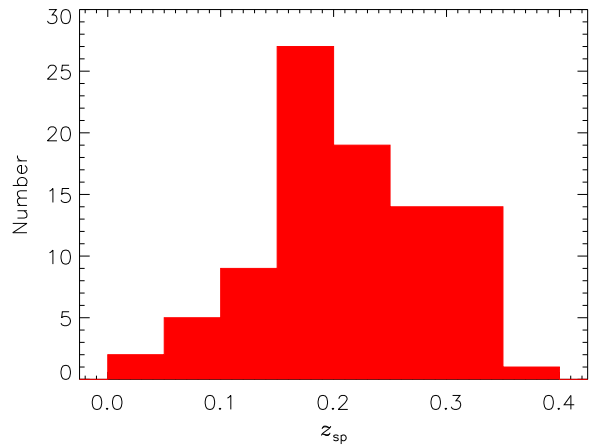}
\end{center}
\caption{Spectroscopic redshift distribution for 91 objects used in to perform emission line diagnostic.}
\label{fig:zhistbpt}
\end{figure}

\indent To produce the BPT diagrams we have measured fluxes for the H$\beta$, [OIII], [NII], H$\alpha$ and [SII] emission lines for 91 objects which have good quality spectra. We show the redshift distribution for these sources in Fig.\ref{fig:zhistbpt}, which are mostly low redshifts ($z<0.4$). The emission line fluxes were measured using the \textsf{splot} routine in IRAF by interactively fitting a Gaussian function to each line. The BPT diagrams, [NII]/H$\alpha$ versus [OIII]/H$\beta$ and [SII]/H$\alpha$ versus [OIII]/H$\beta$ are shown in Fig.\ref{fig:bptdiag}.  The blue line is the pure star formation curve of \cite{kauffmann03}  and the red lines are the extreme starburst lines of \cite{kewley01}. Objects which lie below the blue curve in the [NII]/H$\alpha$ versus [OIII]/H$\beta$ and the red curve in the [SII]/H$\alpha$ versus [OIII]/H$\beta$ are star forming or HII-region galaxies. Composite galaxies lie in the AGN-HII mixing region between the blue and the red curve in the [NII]/H$\alpha$ versus [OIII]/H$\beta$ diagram. The black line in the [SII]/H$\alpha$ versus [OIII]/H$\beta$ diagram is the Seyfert-LINER separation curve of \cite{kewley01}. Objects which lie above the black and red curves are classified as Seyfert and objects that lie above the red curve and below the black line are LINERs. Ambiguous galaxies are those which lie above the extreme star forming line in one diagram and lie below in the other diagram.

\indent Using the emission line diagnostic we found in our sample: 1 Seyfert II galaxy, 34 star forming galaxies, 30 composite galaxies, 5 LINERs and 21 ambiguous galaxies. In addition 13 QSOs or Seyfert I (Type I AGN) were identified from their broad line spectra. We find a large number of ambiguous sources which is most likely due to weak [SII] emission line.

\subsection{Colour diagrams}
\label{sect:ccdiags}
\begin{figure}
\begin{center}
\vspace{0.5cm}
\hspace{0.5cm}
\includegraphics[width=7.7cm, height=7.cm]{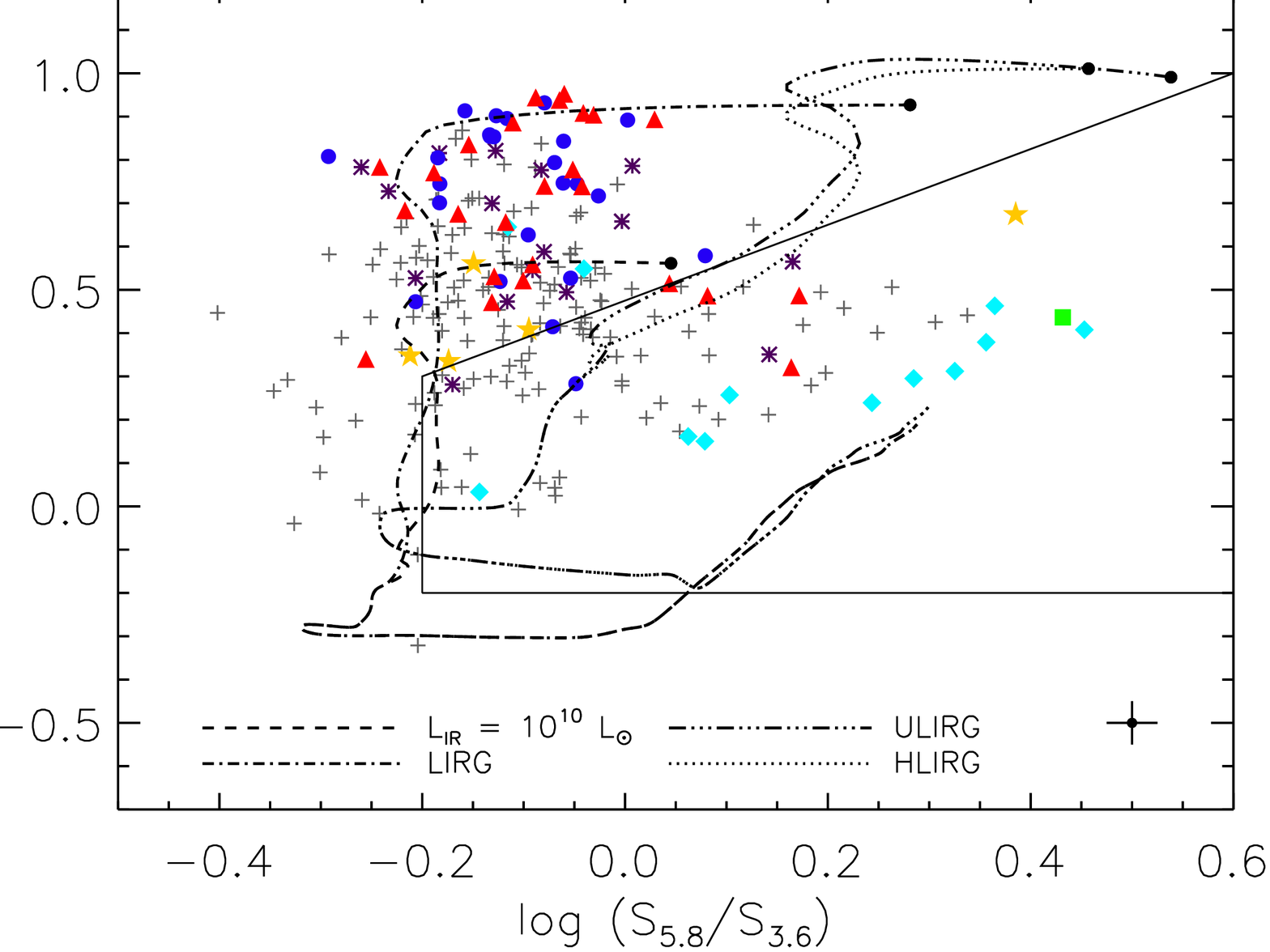}
\end{center}
\caption{IRAC colour-colour plot \citep{lacy04} for 233 70 $\mu$m sources with detection in all IRAC bands. Cyan diamonds are QSOs, red triangles are star forming galaxies, blue circles are composite galaxies, green squares are Seyferts, orange stars are LINERs, pink asterisks are ambiguous galaxies and grey plus signs are galaxies without a spectroscopic classification. The black wedge is the AGN region as defined by \citet{lacy04}. We also plot the colour tracks between $z=0-4$ using the template SEDs of \citet{chary01} for different types of IR galaxies. The black circles represent $z=0$. Typical error bar is shown at the bottom right corner.}
\label{fig:lacyplot}
\end{figure}

We use the SWIRE data with the AF2/WYFFOS spectroscopy to reproduce the \cite{lacy04} IRAC colour-colour diagram. This diagnostic requires the sources to have detections in all four IRAC bands, 3.6, 4.5, 5.8, and 8 $\mu$m. We show in Fig.\ref{fig:lacyplot} the IRAC colour-colour plot for 233 sources in our sample with detections in all the IRAC bands. Objects that lie within the black wedge are expected to be AGN dominated as suggested by \cite{lacy04}. We also plot the colour tracks between $z=0-4$ using the template SEDs of \cite{chary01}.

\indent As expected we find that most of the broad-line objects (10/12) lie within the AGN wedge but we also find that 2 of the broad line AGNs occupy the same parameter space as the star forming galaxies. This is consistent with their IR SEDs which show that the starburst component is the dominant process in the IR. The star forming galaxies occupy the top-left above the AGN wedge which is in agreement with previous work \citep{donley08, trichas09, trichas10}. The narrow-line AGN lie within the AGN wedge. The majority of the LINERs, composite and ambiguous galaxies occupy the same region as the star forming galaxies which implies that these objects are most likely starburst dominated in the NIR. 

\begin{figure*}
\begin{center}
\vspace{1cm}
\hspace{0.3cm}
\includegraphics[height=7.cm, width=8cm]{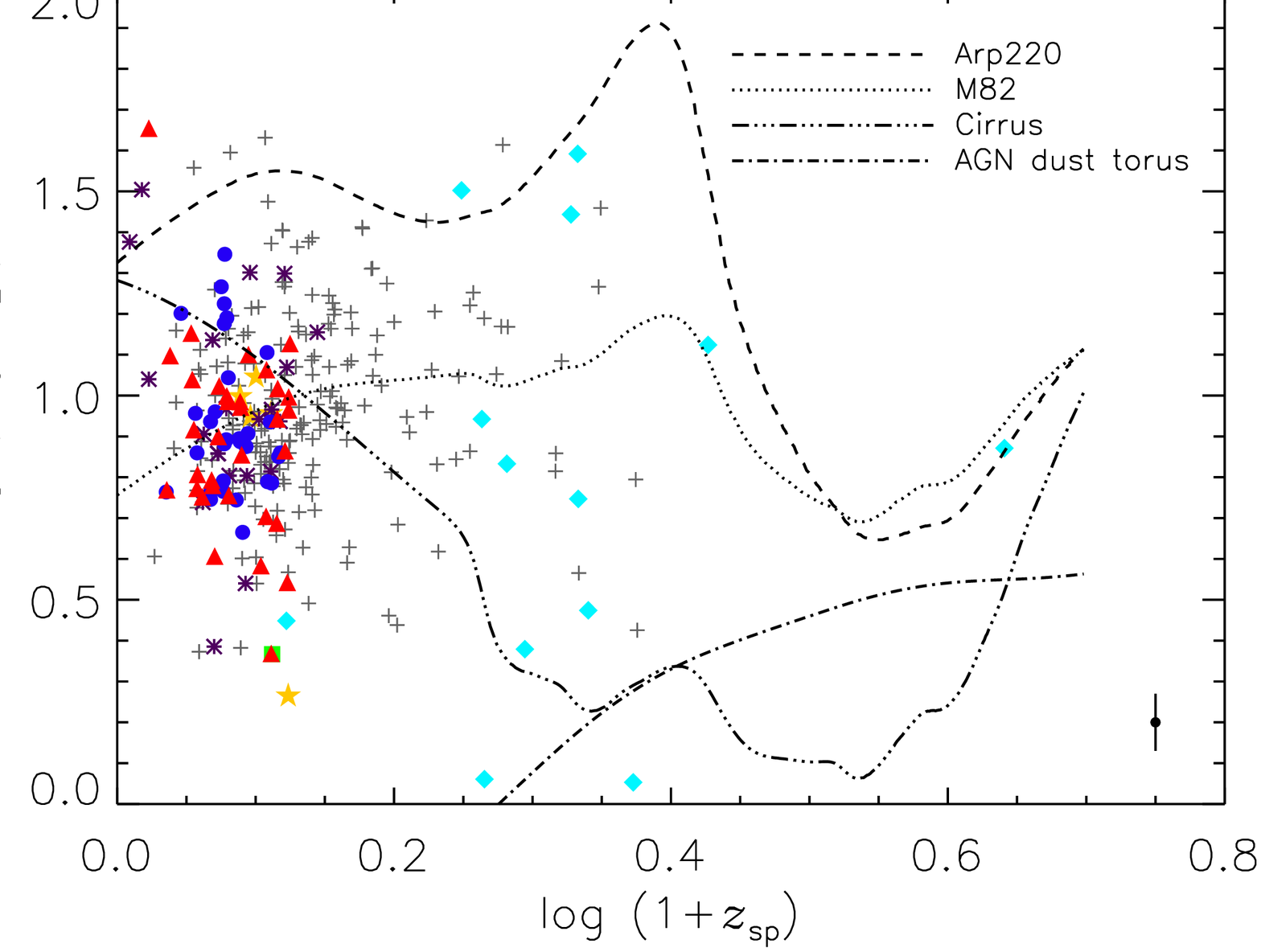}
\hspace{0.8cm}
\includegraphics[height=7cm, width=8cm]{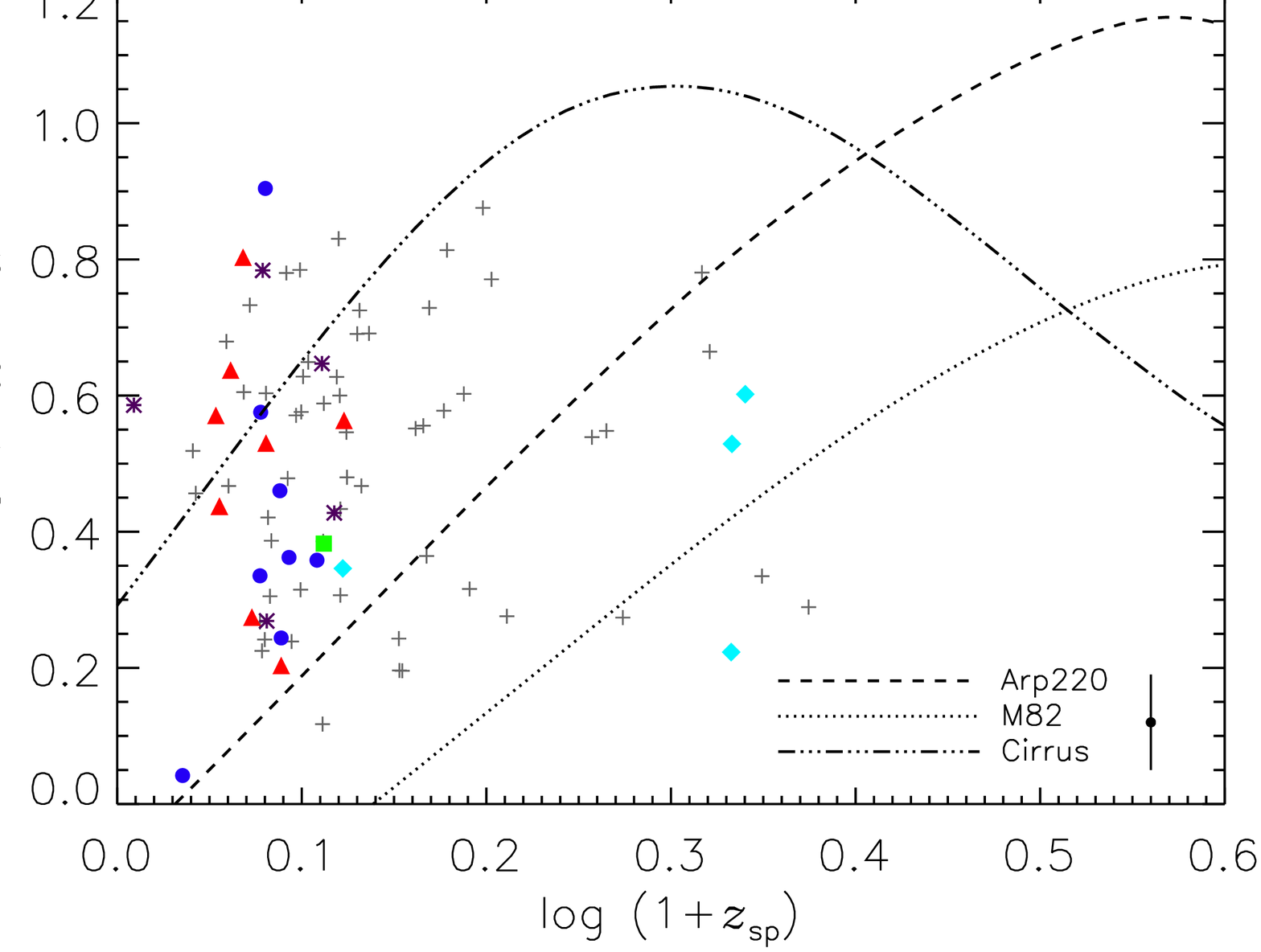}
\end{center}
\caption{\textbf{Left}: Observed 24 and 70 $\mu$m flux ratios as a function of redshift for all 70 $\mu$m sources. Cyan diamonds are QSOs, red triangles are star forming galaxies, blue circles are composite galaxies, green squares are Seyferts, orange triangles are LINERs and grey plus signs are galaxies without a spectroscopic classification. Colour tracks using \citet{mrr04} SED templates. \textbf{Right}: Observed 70 and 160 $\mu$m flux ratios as a function of redshift for 83 70 $\mu$m sources. Purple diamonds are QSOs, red triangles are star forming galaxies, blue circles are composite galaxies, green squares are Seyferts, orange triangles are LINERs and grey plus signs are galaxies without a spectroscopic classification. Colour tracks using \citet{mrr04} SED templates. Note that the AGN dust torus colour track in the right panel lies below the origin and therefore it is not shown. Typical error bars are shown at the bottom right.}
\label{fig:mipscols}
\end{figure*}

\indent In the left panel of Fig.\ref{fig:mipscols}, we plot the ratios of MIPS 24 and 70 $\mu$m flux densities as a function of redshift for all sources in our spectroscopic sample and colour tracks using the SED templates of \cite{mrr04} (The SED models are described in more detail in Sect.\ref{sect:sedfits}). The S$_{70}$/S$_{24}$ ratio can be used to distinguish between star forming galaxies and AGN dominated sources except for sources with strong PAH emission features redshifted into the 24 $\mu$m band (e.g. M82 at $z\sim2$) and for AGNs with strong silicate absorption features at $z\sim1.5$ \citep{frayer06, clements10b}. In this diagnostic, the star forming galaxies are expected to have high S$_{70}$/S$_{24}$ ratio, whereas AGN dominated sources are expected to have low S$_{70}$/S$_{24}$ ratio. The majority of the sources in our sample have observed ratios within the range of values between the Arp220 starburst and cirrus tracks which indicates that a mixture of starburst and cirrus or a pure starburst or cirrus component dominates the FIR emission. 

\indent More recently, \cite{mullaney10} analyse FIR properties of X-ray sources using ultradeep 70 and 24 $\mu$m \textit{Spitzer} observations to show that the 70/24 $\mu$m flux ratio can discriminate an AGN or starburst dominated IR SED for a given AGN. In our sample only 4 of the 13 QSOs appear to have flux ratios similar to AGN dust torus with rest of the QSOs having ratios similar to the starburst component, suggesting that the IR SED is starburst dominated. SED models (see Sect.\ref{sect:sedfits}) of our QSOs shows that most of them have a starburst dominated IR SED.

\indent We show in the right panel of Fig.\ref{fig:mipscols}, the observed 70 and 160 $\mu$m flux ratios for 83 sources with 160 $\mu$m detection, which are a rough estimate of the dust temperature for the FIR peak emission. The S$_{160}$/S$_{70}$ ratio remains roughly constant across the redshift range and is consistent with being cirrus or starburst dominated. S09 and \cite{kartaltepe10} compare their 70 $\mu$m sample with local sources in the S$_{160}$/S$_{70}-z$ plot and find a large fraction of galaxioes that are colder than their local equivalent. Indeed our analysis in Sect.\ref{sect:smgdust} shows that these sources do have dust temperatures that are colder than local IR galaxy population.

\subsection{Spectral energy distribution}
\label{sect:sedfits}
We model the SEDs for our 294 sources following the method described in \cite{mrr05, mrr08} used for the SWIRE photometric redshift catalogue. We have for each source photometry in at least 3 of the 5 optical \textit{U, g, r, i} and \textit{Z} bands and IR photometry from \textit{Spitzer} IRAC 3.6 - 8 $\mu$m and MIPS 24 - 160 $\mu$m bands. The SED fitting follows a two-stage approach, by first fitting the optical to near-IR (U to 4.5 $\mu$m) SED using the six galaxy and three AGN templates used by \cite{mrr08}. We then calculate the IR excess by subtracting the galaxy model fit from the 4.5 to 24 $\mu$m data. We then fit the IR excess, 70 and 160 $\mu$m (for 83 70 $\mu$m sources) data points with the IR template of \cite{mrr04, mrr05, mrr08}. The IR templates are derived from radiative transfer models dependent on interstellar dust grains, the geometry and the density distribution of dust. The IR templates are: 1.) IR 'cirrus`: optically thin emission from interstellar dust illuminated by the interstellar radiation field; 2.) an M82 starburst; 3.) a more extreme Arp220-like starburst and 4.) an AGN dust torus. We also allow the sources to be fit by a mixture of: 1.) M82 starburst and cirrus, 2.) M82 starburst and AGN dust torus and 3.) Arp220 and AGN dust torus to properly represent the IR excess \citep{mrr89, mrr05}.

\begin{figure*}
\begin{center}
\vspace{1.5cm}
\hspace{0.8cm}
\includegraphics[width=16cm, height=15cm]{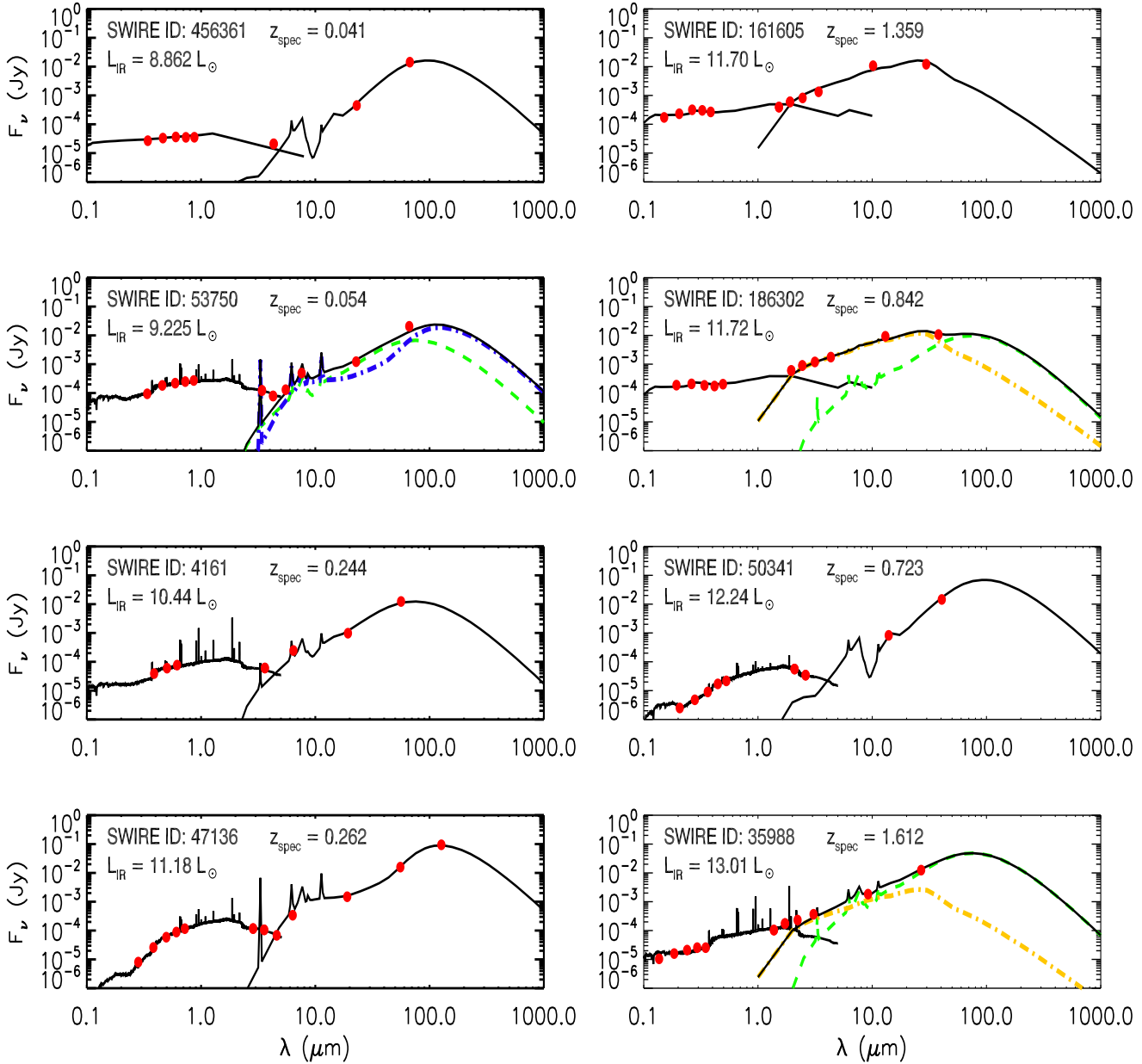}
\end{center}
\caption{SED for a sample of 8 sources at a variety of luminosities. The black lines are the best-fit template SEDs, dashed green line is the M82 starburst, dot-dash blue line is cirrus and orange dot-dash line is AGN dust torus component of the IR SED. All photometry has been redshift-corrected to $z=0$. The errors bars are smaller than the size of the red points.}
\label{fig:mrrseds}
\end{figure*}

\indent We show example SED fits for some of our sources in Fig.\ref{fig:mrrseds}. The IR SED fits suggest that 77 sources require the cold cirrus component, with only 7 fit with a pure cirrus template while 286 sources in our sample require a starburst component. 32 sources require an AGN dust torus component, with only 1 fit with a pure AGN dust torus template, 28 of which occupy the AGN parameter space in Fig.\ref{fig:lacyplot} including 10 spectroscopically identified AGNs (9 QSOs and 1 Seyfert). The SED fits imply that energetically, the 70 $\mu$m sources are powered by star formation and the AGN dust torus makes very little contribution to the IR emission. For sources fit with an AGN dust torus and an M82 starburst mixture, the MIR emission is AGN dominated with the FIR emission powered by the starburst component. This finding is to be expected as our selection at 70 $\mu$m directly implies that the FIR emission is dominated by star formation as opposed to AGN which peak at much shorter wavelengths (NIR/MIR) and are weak FIR emitters \citep{alonsoherrero06}.

\indent For the 14 AGNs (13 QSOs and 1 Seyfert) in our sample we find 7 (6 QSOs and 1 Seyfert) are fit by a mixture of AGN dust torus and M82 starburst, 2 QSOs fit with a mixture of Arp220 starburst and AGN dust torus, 1 QSO fit by pure AGN dust torus, 2 QSOs fit with M82 starburst template and 2 QSOs are fit with an Arp220 starburst SED i.e. no AGN dust torus. The AGN dust torus component in these sources typically contributes $\sim$ 40 - 100\% to the total IR emissions. All galaxies classified as LINERs are fit with the M82 starburst template. Only 1 of the composite galaxies requires the AGN dust torus component, while the rest are fit with a pure starburst or a mixture of M82 and cirrus. Of the 21 ambiguous galaxies, 2 are fit with a mixture of AGN dust torus and M82 starburst whilst the rest are fit with a pure starburst or M82 starburst and cirrus mixture. 3 of the galaxies identified as star forming from the BPT diagrams require an AGN dust torus component, while only 1 is Arp220 like starburst with the rest fit with a pure M82 starburst or M82 and cirrus mixture. 12 of the unclassified sources require the AGN dust torus component to model the IR emission. 

\begin{table*}
\begin{center}
\begin{tabular}{cccccccc}
\hline
Optical Classification & \multicolumn{7}{c}{IR SED type}\\
& Arp220 & M82 & Cirrus & AGN DT & M82 and Cirrus & M82 and AGN DT & Arp220 and AGN DT\\
\hline
QSOs & 2 & 2 & - & 1 & - & 6 & 2\\
Seyferts & - & - & - & - & - & 1 & -\\
Star-forming & 1 & 15 & 1 & - & 13 & 3 & -\\
Composite & 5 & 12 & 2 & - & 10 & 1 & -\\
LINERs & - & 5 & - & - & - & - & -\\
Ambiguous & 5 & 7 & 2 & - & 7 & - & - \\
None & 44 & 86 & 4 & - & 40 & 15 & 1\\
\hline
\end{tabular}
\end{center}
\label{tab:sedfitsummary}
\caption{Summary of the number of objects fit with the 7 IR SEDs separated by optical classification.}
\end{table*}

\indent We also have in our sample 83 70 $\mu$m sources that have a 160 $\mu$m detection, with IR luminosities in the range $\sim$ 10$^{8.6}$ - 10$^{13.7}$ L$_\odot$ and redshift range 0.021-1.369. Of these sources 49 are fit with a cool cirrus component and have typical luminosities of $\sim$ 10$^{11}$ L$_\odot$. This shows that a large fraction of the 70 $\mu$m sources with 160 $\mu$m detection have significant cold dust that contributes $\sim$ 40 - 100\% to the total IR luminosity. Using SED fitting S09 found that the majority of the 70 $\mu$m sources in their sample peak at longer wavelengths ($\lambda > $ 90 $\mu$m) which they interpret as evidence for evolution in the cold cirrus component from $z\sim0$ to higher redshifts. However, at high redshift the 160 $\mu$m data point probes shorter wavelength (80 $\mu$m at $z=1$) and therefore the peak of the emission is not well constrained and a wide range of SED templates can fit the data points. In order to properly understand the role of cold dust we require observations across a wider range of FIR and submm wavelengths which the Herschel Space Observatory \citep{pilbratt10}  is well placed to do. In fact \cite{mrr10} have shown the need for cirrus templates with colder (T = 10 - 20K) dust to model the SEDs of \textit{Herschel} sources. In Sect.\ref{sect:dustprops}, we fit single temperature modified blackbody SEDs to these sources and study in more detail the link between 70 $\mu$m sources and SMGs.

\begin{figure}
\begin{center}
\vspace{0.5cm}
\hspace{0.4cm}
\includegraphics[width=7.8cm, height=7cm]{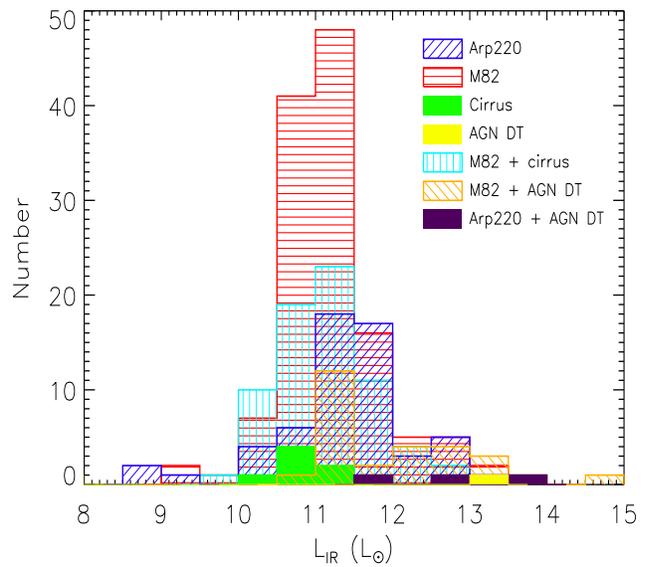}
\end{center}
\caption{Histogram of L$_\mathrm{IR}$ separated by the best fit IR template. Red = m82 starburst only, cyan = M82 and cirrus mixture and black = AGN dust torus only100, green = cirrus only,  blue = Arp220 starburst only, orange = M82 and AGN dust torus mixture and purple Arp220 and AGN dust torus mixture.}
\label{fig:lirhist}
\end{figure}

\indent We summarise the results of the SED fitting in Table \ref{tab:sedfitsummary} and show in Fig.\ref{fig:lirhist} the histogram of the total IR luminosity (calculated by integrating the SED models from 8 to 1000 $\mu$m) separated by the best fitting IR template. Overall, $\sim$ 96\% of the sample requires a starburst component, with $\sim$ 76\% represented by M82 and $\sim$ 20\% by Arp220-like respectively. The M82 starburst template peaks at a shorter wavelength in the FIR and has a less prominent silicate absorption feature indicating less extreme obscuration towards the starburst region. The Arp220-like template on the other hand is characterised by a more prominent silicate absorption feature due to the higher optical depth resulting in larger obscuration. Moreover, the emission peaks at longer wavelengths because an additional cold cirrus component is needed to match the submm emissions. As noted above, significant fraction of 70 $\mu$m sources with 160 $\mu$m detections are fit with the cirrus template (either as a mixture or pure) which is needed to account for the IR emission longword of 100 $\mu$m. Our results suggests that in general the IR emissions in the 70 $\mu$m population is largely powered by star formation, with the cold cirrus component making a significant contribution.

\begin{figure*}
\begin{center}
\vspace{0.5cm}
\hspace{0.2cm}
\includegraphics[width=7.8cm, height=7cm]{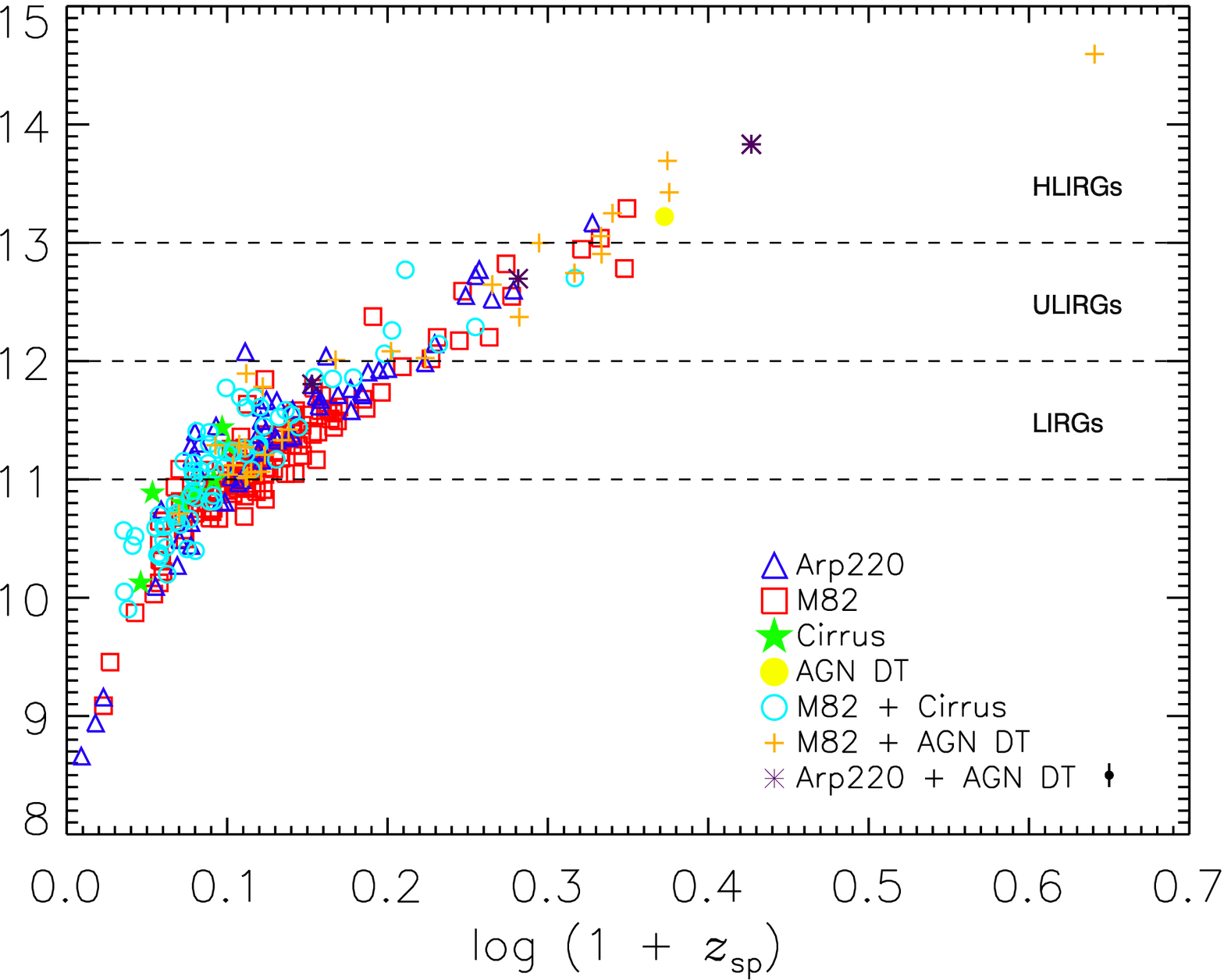}
\hspace{0.7cm}
\includegraphics[width=7.8cm, height=7cm]{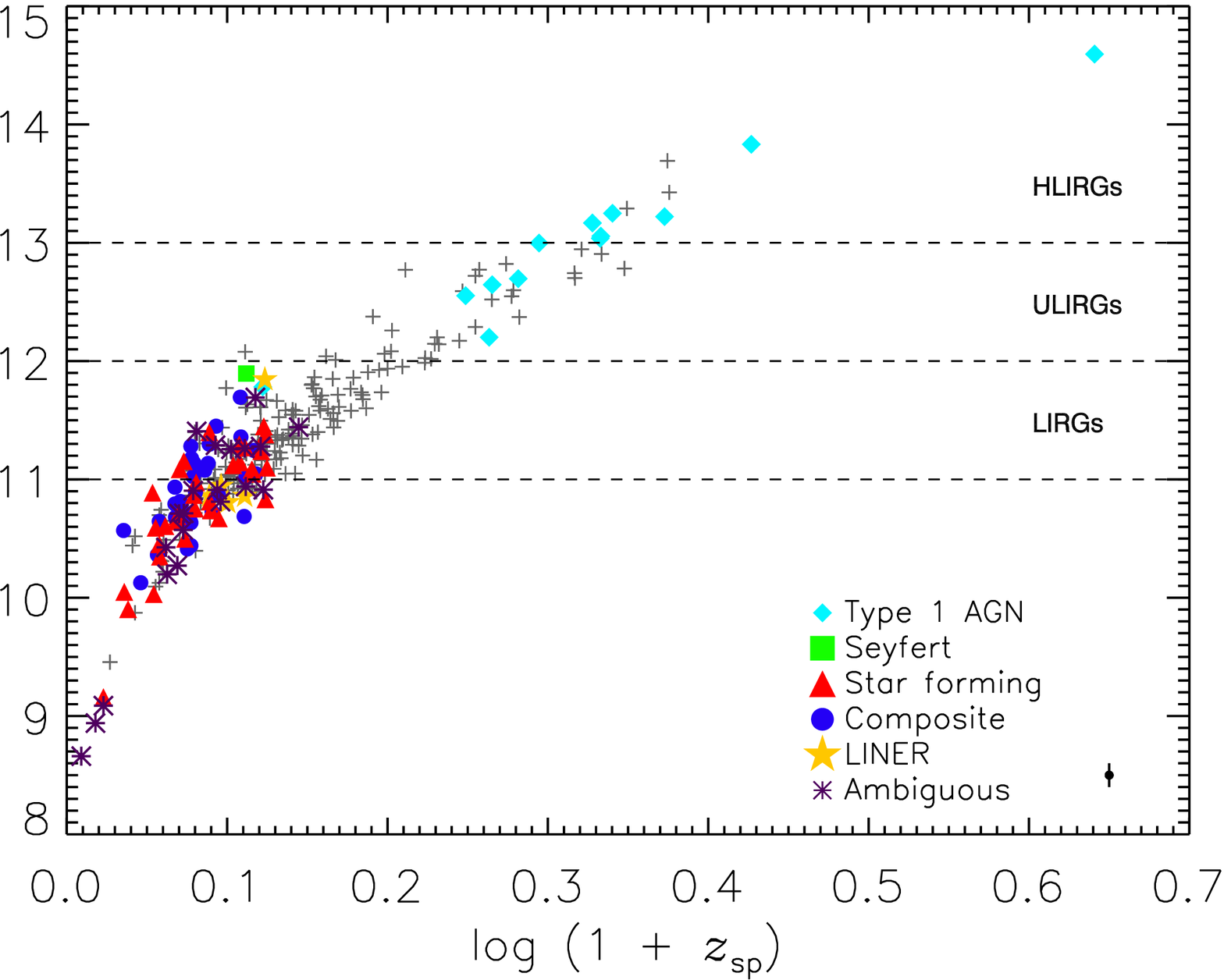}
\end{center}
\caption{\textbf{Left}: Total IR luminosity (L$_\mathrm{IR}$) as a function of redshift for all 293 sources in SWIRE XMM-LSS and LH-ROSAT coded by best fit IR SED type. \textbf{Right}: Total IR luminosity (L$_\mathrm{IR}$) as a function of redshift coded by spectral type. Typical error bar of 0.1 dex is shown at the bottom right.}
\label{fig:lirvz}
\end{figure*}

\indent We show in the left panel of Fig.\ref{fig:lirvz},  L$_\mathrm{IR}$ against redshift colour coded by the best fitting IR template. Our sample consists of 7 normal galaxies defined as L$_\mathrm{IR} < $ 10$^{10}$ L$_\odot$, 93 starburst galaxies (L$_\mathrm{IR}$ = 10$^{10}$ - 10$^{11}$ L$_\odot$), 151 LIRGs, 33 ULIRGs and 10 HLIRGs. We find that the fraction of sources requiring an AGN dust torus component in the SED models increases with the total IR luminosity from 1.1\% at L$_\mathrm{IR} < $ 10$^{10}$ to 10\% for LIRGs,  27.3\% at ULIRGs and 70\% at HLIRGs. This trend is also seen with the spectroscopically identified AGNs where we find 2 LIRGs, 5 ULIRGs and 7 HLIRGs. This finding is consistent with the study of \cite{veilleux95} and \cite{veilleux99}, whose study of IRAS selected galaxies shows that the AGN fraction increases with L$_\mathrm{IR}$.  Our results are in agreement with \cite{kartaltepe10} who find in their sample of 70 $\mu$m sources that the AGN fraction increases from 2\% for L$_\mathrm{IR} < $ 10$^{10}$ L$_\odot$, to 10\% for LIRGs, 51\% for ULIRGs and 97\% for HLIRGs. Similarly \cite{symeonidis10} find in their 70 $\mu$m sample, AGN fractions of 0\% for L$_\mathrm{IR} < $ 10$^{10}$ L$_\odot$, 11\% for LIRGs and 23\% for ULIRGs, being consistent with our results. 

\indent In the right panel of Fig.\ref{fig:lirvz}, we show the distribution of total IR luminosity as a function of redshift coded by optical spectral type determined in Sect.\ref{sect:bpt}. Star forming and composite galaxies span a range of IR luminosities, $\sim$ 10$^9$ - 10$^{12}$ L$_\odot$, with most classified as starburst (L$_\mathrm{IR}$ = 10$^{10}$ - 10$^{11}$ L$_\odot$) or LIRGs. Ambiguous galaxies also span a wide range of IR luminosities, $\sim$ 10$^{8.5}$ - 10$^{12}$ L$_\odot$. Most of the LINERs are seen near the starburst and LIRG classification line. The only Seyfert II object is a LIRG while Type I AGNs are the most luminous sources in the sample and are typically ULIRGs and HLIRGs. Our results are consistent with previous studies, which suggests AGNs are more likely to be found in the most luminous of the IR galaxy population. We note that the source at z = 3.374, which has an extreme IR luminosity (L$_\mathrm{IR}$ = 14.6 L$_\odot$) is consistent with the source being a QSO. However, the derived IR luminosity may be unreliable as the FIR SED of this source is not well constrained since at z = 3.374, the 70 $\mu$m band corresponds to $\sim$ 15 $\mu$m in the rest frame. We will be able to obtain a better estimate of L$_\mathrm{IR}$ for this and other sources by constructing full SEDs from the optical to submm using observations from the Herschel Multi-tiered Extragalactic Survey (HerMES) programme (Oliver et al in prep.).

\subsection{Dust properties}
\label{sect:dustprops}

In this section we select the 70 $\mu$m sources with 160 $\mu$m detection at $z<1.2$ (to ensure we are still sampling the peak of the FIR SED, which at $z=1.2$ 160 $\mu$m corresponds to$\sim$ 73 $\mu$m) to model the FIR SED for 81 objects by a standard single temperature modified blackbody given by:

\begin{equation}
F_\nu = \nu^\beta{B}_\nu(\nu,T_d)
\label{eq:greybody}
\end{equation}

\noindent where \textit{T$_d$} is the dust temperature and $\beta$ is the dust emissivity, corresponding to the slope of the Rayleigh-Jeans tail. This model is a simple approach as the parameters that define the SED do not account for the detailed geometrical mix of dust grains at different temperatures in the interstellar medium (ISM). Nevertheless several studies \citep{hildebrand83, dunne00, dunne01, klaas01, bendo03, blain03, vlahakis05, clements10} have used this method to fit the FIR to submm SED.

\indent We use the same method used for the study of SCUBA Local Universe Galaxy Survey (SLUGS) sources \citep{dunne00, vlahakis05} and local ULIRGs \citep{clements10} by calculating the dust temperature using a $\chi^2$ minimisation method and fixing the emissivity parameter, $\beta$ to 1.5 \citep{chapman05, frayer06, kovacs06, coppin08}.  We also calculate the dust masses in these objects using:

\begin{equation}
M_d = \frac{S_{160}D^2_L}{\kappa_d(\nu)B(\nu, T_d)}
\label{eq:mdust}
\end{equation}

\noindent where S$_{160}$ is the observed 160 $\mu$m flux, D$_L$ is the luminosity distance to the object and $\kappa_d$ is the dust opacity coefficient. The value of $\kappa_d$ is 1.139 m$^2$ kg$^{-1}$ at 160 $\mu$m is interpolated from \cite{draine03}. Note that our derived dust masses are crude estimates because of the uncertainties in $\kappa_d(\nu)$ and the dust temperatures, resulting from the complexity of the properties of interstellar dust (see review by \citealt{draine03}). 

\begin{figure*}
\begin{center}
\vspace{0.5cm}
\includegraphics[width=7.5cm,height=5.5cm]{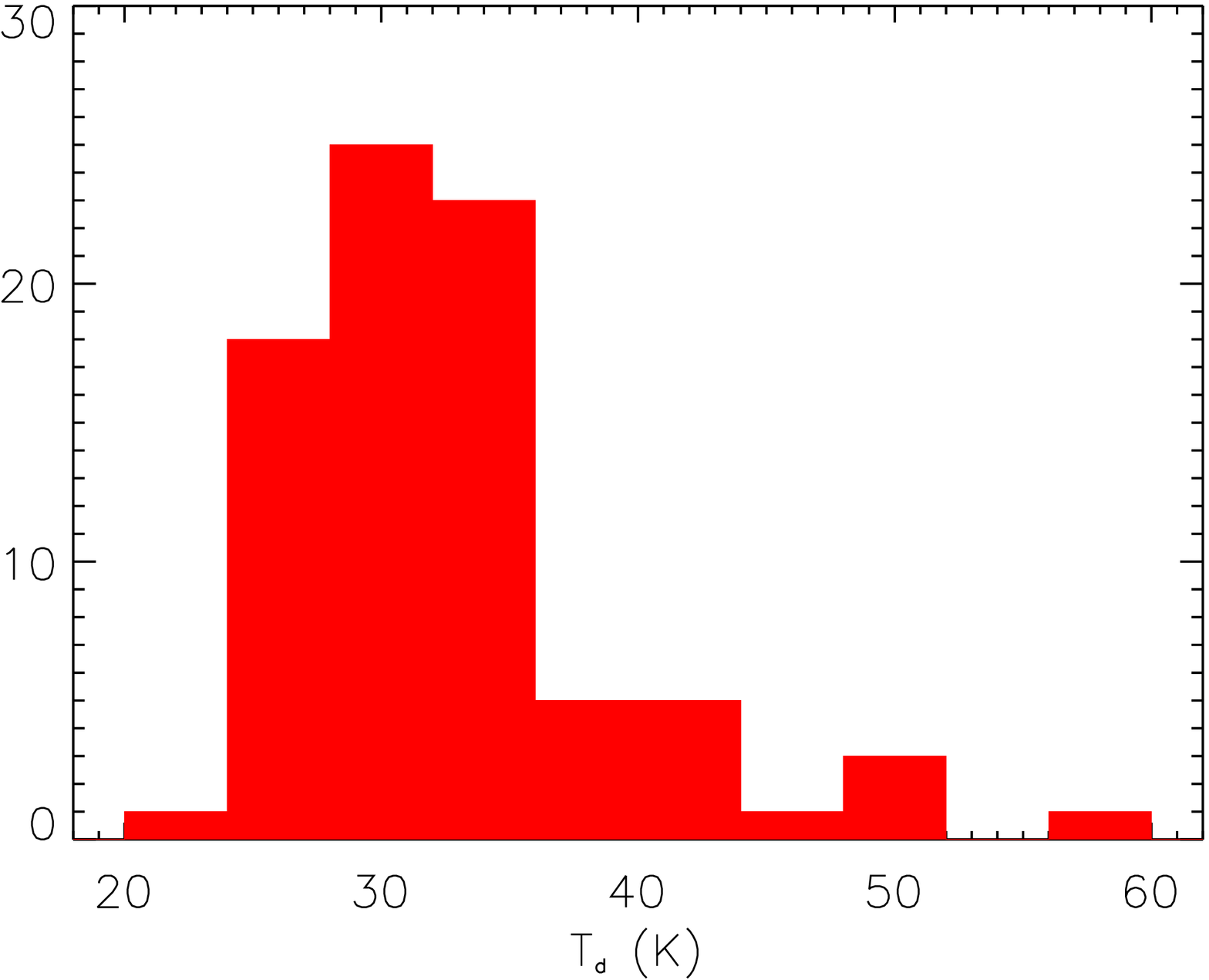}
\hspace{0.5cm}
\includegraphics[width=7.5cm,height=5.5cm]{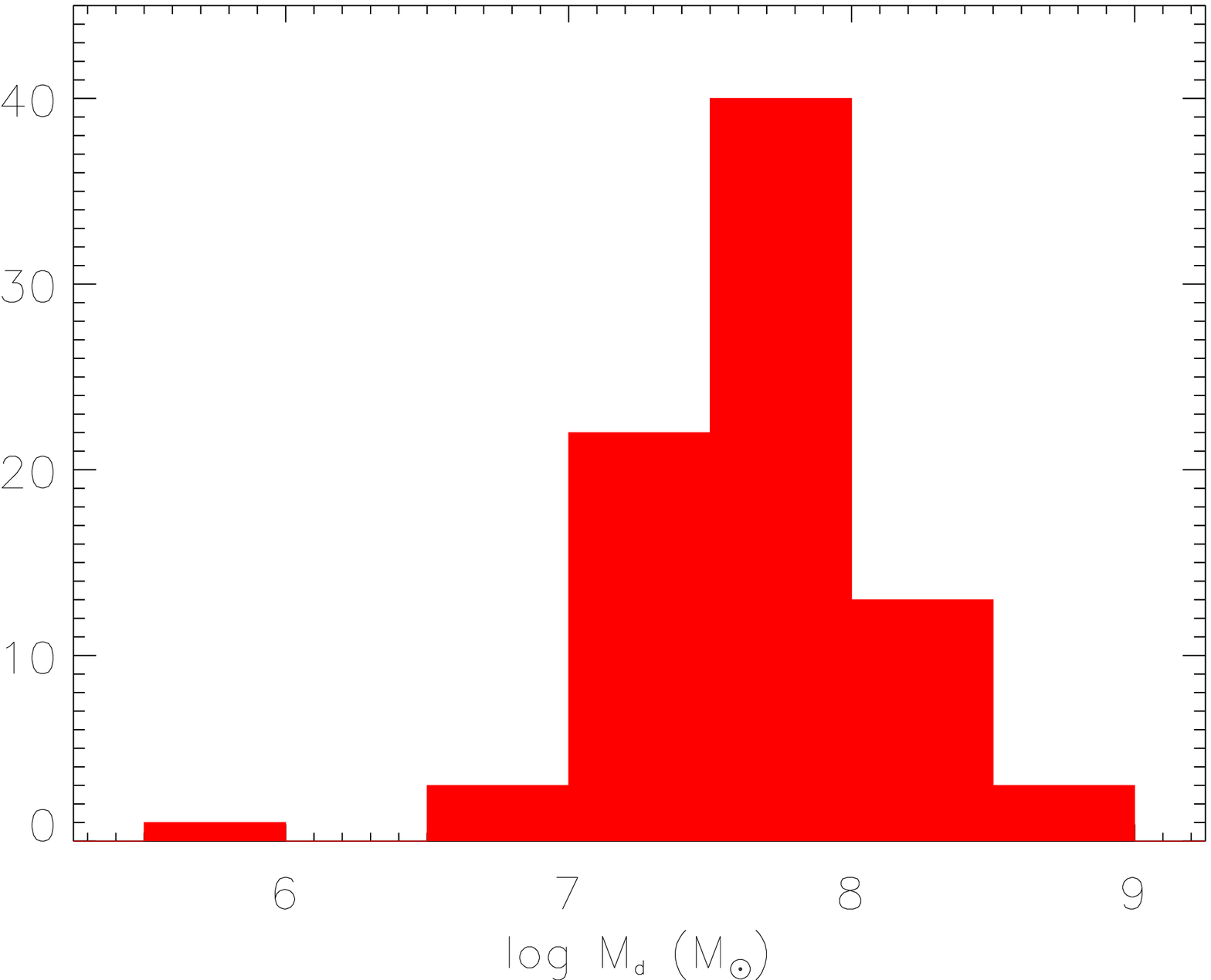}
\end{center}
\caption{\textbf{Left}: Dust temperature distribution estimated using Eq.\ref{eq:greybody}. \textbf{Right}: Dust mass distribution estimated from Eq.\ref{eq:mdust}.}
\label{fig:dustprops}
\end{figure*}

\indent Our sample of 82 sources is characterised with mean and standard deviation for the best fitting dust temperatures, T$_d$ = 32.7 $\pm$ 6.7K and dust mass range, M$_d$ $\sim$ 10$^5.7$ - 10$^8.8$ M$_\odot$ for $\beta$ = 1.5. Our dust temperature and dust mass distribution shown in Fig.\ref{fig:dustprops} are consistent with previous studies. \cite{frayer06} estimate dust temperatures, T$_d \simeq$ 30 $\pm$ 5 K for 70 and 160 $\mu$m detected sources from the \textit{Spitzer} extragalactic First Look Survey (xFLS) for $\beta$ = 1.5. \cite{dunne00} SLUGS sample is characterised by dust temperatures, T$_d$ = 35.6 $\pm$ 4.9 and dust mass range, M$_d$ = 10$^6$ - 10$^9$ M$_\odot$. \cite{dunne00} also fit the emissivity parameter finding $\beta$ = 1.3 $\pm$ 0.2. If we adopt $\beta$ = 1.3, we determine slightly warmer dust temperatures, T$_d$ = 33.9 $\pm$ 7K. 

\indent In our analysis we have assumed the dust emissivity to be 1.5 but other studies \citep{dunne01, klaas01, vlahakis05} have shown that the FIR-submm SED may be better represented by two dust component fits. This model consists of dust at two different temperatures, a cold (T$_d \leq$ 25K) and a warm component  (T$_d > $ 30K) and the emissivity parameter fixed at 2. However these studies have been carried out on local sources for which they have observations across a wide spectral range from FIR (IRAS, ISO and \textit{Spitzer} photometry) to submm (SCUBA photometry). For our sources however, the MIPS 160 $\mu$m provides the longest data point and therefore it's not possible to fit a multicomponent dust SED to our sources. Even constraining the dust emissivity parameter, which varies between 1-2, is not possible as this requires data points across a range of wavelengths from the FIR to submm.

\indent By fixing $\beta$ = 2, our sample is characterised by T$_d$ = 30 $\pm$ 6 K, so increasing the emissivity results in slightly lower dust temperatures. Recent study by \cite{planckcolabclements11} of \textit{Planck} sources has been crucial in resolving the degeneracy between $\beta$ and T$_d$ for nearby galaxies. Their results confirm cold dust is seen in \textit{Planck} sources, which have median T$_d$ = 26.3 K with temperatures ranging from 15 - 50 K and $\beta$ = 1.2. \cite{amblard2010aa5189} find for \textit{Herschel}-ATLAS sources $\beta$ = 1.4 $\pm$ 0.1 which is consistent with $\beta$ = 1.3 for SLUGS sources \citep{dunne00}. Thus the \textit{Herschel} Space Observatory (\textit{Herschel}) and the \textit{Planck} Surveyor will play a crucial role in accurately determining the Rayleigh-Jeans tail of the dust SED. The SWIRE XMM-LSS and LH fields have been observed as part of the HerMES programme, which will allow us to accurately determine the shape of the Rayleigh-Jeans tail of the dust SED and provide an insight into the role of cold dust. 

\subsubsection{Link to SMGs}
\label{sect:smgdust}

\begin{figure}
\begin{center}
\vspace{0.4cm}
\hspace{1.cm}
\includegraphics[width=8.cm, height=7.cm]{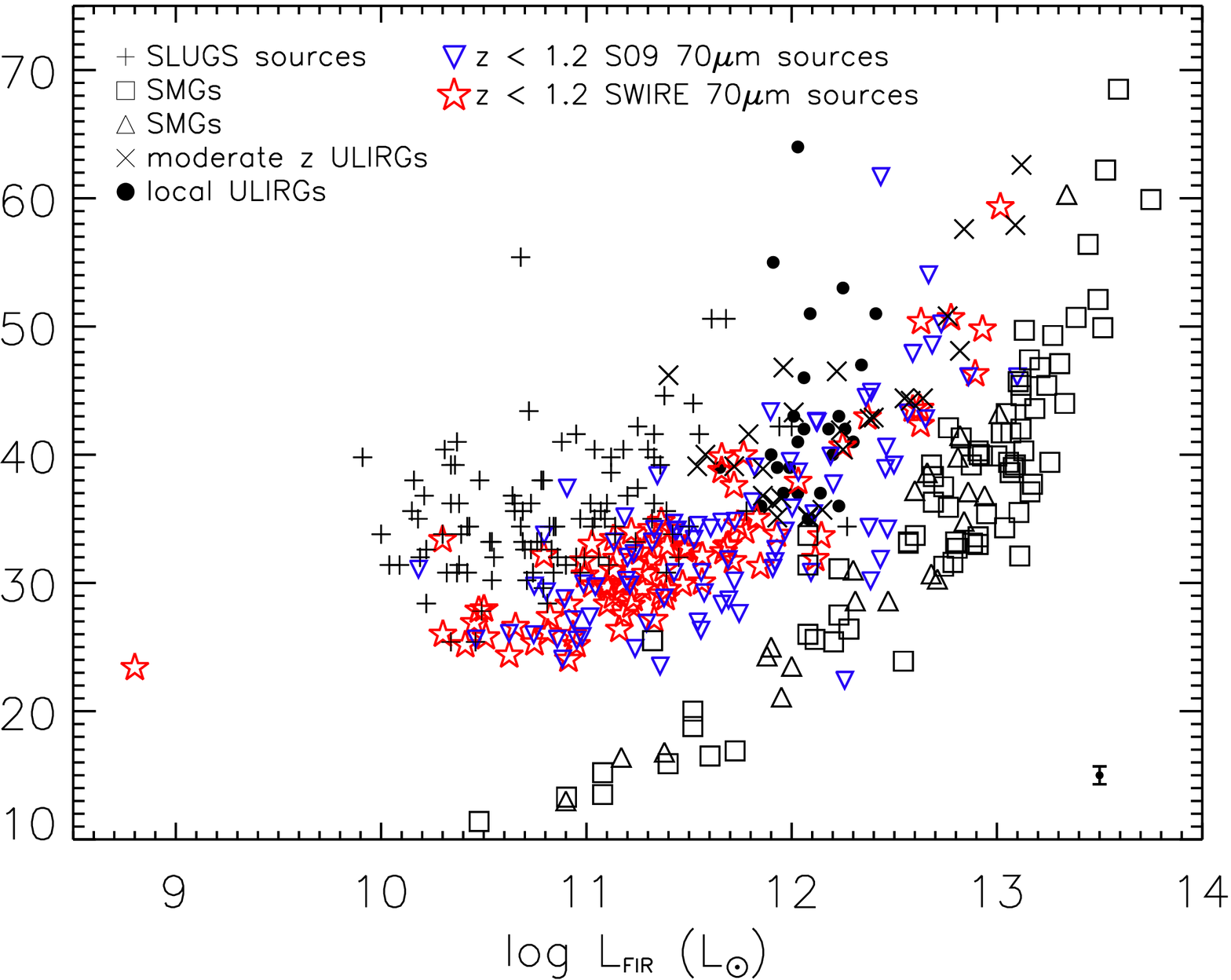}
\end{center}
\caption{Temperature-luminosity diagram comparing ULIRGs (crosses: \citealt{yang07}; filled circles: \citealt{clements10}), SMGs (squares: \citealt{chapman05}; triangles: \citealt{kovacs06}), normal SLUGS galaxies \citep{dunne00} and S09 70 $\mu$m sample based on single temperature dust SED fits, with $\beta$ = 1.5. L$_\mathrm{FIR}$ is derived by integrating Eq.\ref{eq:greybody} between 40 - 1000 $\mu$m. Diagram adapted from \citet{clements10}. Typical error bars are shown at the bottom right.}
\label{fig:ltplot}
\end{figure}

\indent In this section we plot our 82 70 $\mu$m sources on the luminosity-temperature (L-T) diagram of \cite{yang07} to test the link between local and intermediate redshift galaxies to the high-redshift SMGs. The luminosities are estimated by integrating Eq.\ref{eq:greybody} between 40 and 1000 $\mu$m which we define as L$_\mathrm{FIR}$. The two galaxy populations appear to lie in different parts of the L-T plane such that the SMGs appear to have colder dust temperatures and higher dust masses which is seen as evidence for evolution in the dust properties of galaxies. In Fig.\ref{fig:ltplot} we plot the positions of our 70 $\mu$m sources alongside SMGs from \cite{chapman05, coppin08} and \cite{kovacs06} as well as moderate-$z$ ULIRGs from \cite{yang07}, more normal SLUGS sources from \cite{dunne00}, local ULIRGs from \cite{clements10}. We also plot the 70 $\mu$m sample of S09 selected at $z<1.2$, S$_{70} > $ 9mJy and S$_{160} > $ 50mJy by estimating T$_d$ and L$_\mathrm{FIR}$ using our method. The main difference between the two 70 $\mu$m samples is that our work is restricted to sources with spectroscopic redshifts while some of the sources in the S09 sample have photometric redshift estimates. Furthermore, only 62\% of the S09 sample has $> 5\sigma$ detection at 160 $\mu$m. Thus the error estimates for the S09 sample in T$_d$ and L$_\mathrm{FIR}$ are expected to be larger.

\indent We find that our 70 $\mu$m sources (red stars) in Fig.\ref{fig:ltplot} lie systematically below the plane of the SLUGS sources (plus sign) and local and moderate-$z$ ULIRGs (crosses and filled circles) in the L-T diagram. Our 70 $\mu$m sources span a range of temperatures $\sim$ 24 - 60 K with $\sim$ 30\% having T$_d \leq$ 30 K, whereas only $\sim$ 6\% of the SLUGS sample and none of the moderate and local ULIRGs have T$_d \leq$ 30 K. This implies that our 70 $\mu$m sources are noticeably cooler than what has been observed previously in the local to intermediate IR luminous sources. The 70 $\mu$m sample of S09 is characterised by dust temperature T$_d$ = 34.3 $\pm$ 7.45 K, where $\sim$ 31\% have T$_d <$ 30K, which implies that the dust properties of the two 70 $\mu$m samples are similar. Thus it appears that the 70 $\mu$m population is beginning to fill the gap between the cold high-$z$ SMGs and the warmer local to moderate-$z$ IR galaxies.

\indent The detection of cooler dust can be interpreted as the 70 $\mu$m sources representing a missing link between the two population of galaxies \citep{symeonidis09} or as a result of selection effect \citep{yang07}. The studies of the local to moderate-$z$ IR galaxies are based on IRAS 60 $\mu$m derived samples and therefore these studies may have been biased to warmer dust temperatures whereas the high-$z$ SMGs are may be biased towards cold dust temperatures. Thus using observations at the longer 70 $\mu$m band, we are most likely observing galaxies that have cooler dust temperatures when compared to the IRAS selected sample and therefore bridging the separation between the high-$z$ SMGs and local sources. 

\indent Results from a \textit{Planck} study of nearby galaxies \citep{planckcolabclements11} has found evidence for cold dust at T $<$ 20K with a variable emissivity and therefore reducing the distinction between the two populations. The \textit{Planck} sources overlap with the SLUGS galaxies but extend to colder temperatures with temperatures ranging from 15 - 50K. Studies carried out by \cite{amblard2010aa5189} using \textit{Herschel} observations have also begun to fill in the gap between local IRAS galaxies and high-$z$ SMGs. \cite{planckcolabclements11} conclude that cold dust is significant and remains an unexplored component in many nearby galaxies. As mentioned in the previous section we will be able to further understand the nature of the 70 $\mu$m sources using observations from \textit{Herschel}, which will provide data points across wide range of wavelengths from 70 - 500 $\mu$m.

\section{Summary and Conclusions}
\label{sect:summary}
We present spectroscopic follow-up observations of 70 $\mu$m selected sources in the SWIRE XMM-LSS and LH-ROSAT regions using the multi-object fibre spectrograph AF2/WYFFOS  on WHT. We measured new spectroscopic redshifts for 293 70 $\mu$m and 35 24 $\mu$m sources. The redshift distribution for the 70 $\mu$m sources peaks at $z\sim0.3$ and has a high redshift tail out to $z\sim3.5$. The majority of the 24 $\mu$m selected QSOs are at typically $z>1$. The spectroscopic redshifts of our sample were compared with the photometric redshift estimates which show a good agreement for galaxies while some QSOs have poorer agreement which we attribute to optical variability or photometric redshift aliasing.

\indent We carry out emission line diagnostic for 91 70 $\mu$m sources with [OIII], H$\beta$, [NII], H$\alpha$ and [SII] emission lines present to classify the galaxies into star-forming, Seyfert, composite and LINERs. We find in our sample 34 star-forming, 30 composite, 1 Seyfert, 5 LINERs and 21 ambiguous galaxies and 13 QSOs identified from their broad emission line spectra. 


\indent We then modelled the SEDs for each source from optical to the FIR using the method described in \cite{mrr05, mrr08} and SED templates from \cite{mrr08} . Our major finding is that the IR emission in the 70 $\mu$m sources are powered by star formation, with the AGN dust torus component making a small contribution for non QSOs. \cite{symeonidis10} examine the AGN content of a sample of IR luminous 70 $\mu$m selected galaxies by modelling the panchromatic SEDs from X-ray to IR and find for most of their sources, the AGN contributes less than 10\% to the IR budget. Moreover, they find all sources in their sample are primarily powered by star-formation. Our results are also consistent with the findings of \cite{trichas09} who find their X-ray selected sample with 70 $\mu$m detection are strongly star-forming and require a starburst SED to fit the 70 $\mu$m and 160 $\mu$m (if available) photometry. For 9 QSOs and the 1 Seyfert galaxy the AGN dust torus dominates in the NIR and MIR regimes producing from 40 to 100\% of the total IR luminosity. Of the 83 70 $\mu$m sources in our sample with 160 $\mu$m detections, 49 require the cirrus component which implies the presence of large amounts of cold dust at T$_d \sim$ 25 K. By integrating the SEDs from 8 - 1000 $\mu$m, we find our 70 $\mu$m sample is dominated by starbursts and LIRGs. We also find that the AGN fraction increases with L$_\mathrm{IR}$ with almost all (7/10) HLIRGs being optically identified as QSOs.

\indent Finally, we fit single temperature modified blackbodies to 82 70 $\mu$m sources with 160 $\mu$m fluxes at $z<1.2$ with $\beta$ fixed at 1.5. Our sample has a mean and standard deviation dust temperatures, T$_{d}$ = 32.7 $\pm$ 6.7 K and dust mass range, M$_{d}$ = 10$^6$ - 10$^9$ M$_\odot$ and even increasing $\beta$ to 2 results in slightly lower temperatures, T$_{d}$ =  30 $\pm$ 6 K. In general we find that as  $\beta$ increases T$_{d}$ decreases. Examining our sources in the L-T diagram shows that the 70 $\mu$m sources with 160 $\mu$m detection have dust temperatures that are systematically lower than the local IR luminous galaxies. \textit{Herschel} and \textit{Planck} will play a vital role in characterising the IR galaxy population, as well as breaking the degeneracy between $\beta$ and T$_{d}$ by providing data points across range of FIR/submm wavelengths.

\section{Acknowledgements}
HP acknowledges financial support from STFC and the support staff at WHT. The authors would like to thank Myrto Symeonidis for useful discussions and comments. We also thank Alberto Franceschini, Antonis Georgakakis, Lucia Marchetti, Ismael P{\'e}rez-Fournon, Nieves Castro-Rodr{\'{\i}}guez, and Markos Trichas. This work is based on observations made with the \textit{Spitzer} Space Telescope, which is operated for NASA by the Jet Propulsion Laboratory, California Institute of Technology.

\bibliographystyle{mn2e}
\bibliography{swire_spectroscopy}

\begin{thebibliography}{67}
\expandafter\ifx\csname natexlab\endcsname\relax\def\natexlab#1{#1}\fi

\bibitem[{{Alonso-Herrero} {et~al.}(2006){Alonso-Herrero},
  {P{\'e}rez-Gonz{\'a}lez}, {Alexander}, \& {et al.}}]{alonsoherrero06}
{Alonso-Herrero} A., {P{\'e}rez-Gonz{\'a}lez} P.~G., {Alexander}, {et al.},
  2006, \apj, 640, 167

\bibitem[{{Amblard} {et~al.}(2010){Amblard}, {Cooray}, {Serra}, \& {et
  al.}}]{amblard2010aa5189}
{Amblard} A., {Cooray} A., {Serra} P., {et al.}, 2010, \aap, 518, L9+

\bibitem[{{Babbedge} {et~al.}(2004){Babbedge}, {Rowan-Robinson},
  {Gonzalez-Solares}, \& {et al.}}]{babbedge04}
{Babbedge} T.~S.~R., {Rowan-Robinson} M., {Gonzalez-Solares} E., {et al.},
  2004, \mnras, 353, 654

\bibitem[{{Baldwin} {et~al.}(1981){Baldwin}, {Phillips}, \&
  {Terlevich}}]{baldwin81}
{Baldwin} J.~A., {Phillips} M.~M., {Terlevich} R., 1981, \pasp, 93, 5

\bibitem[{{Bendo} {et~al.}(2003){Bendo}, , {Joseph}, {Wells}, \& {et
  al.}}]{bendo03}
{Bendo} G.~J., , {Joseph} R.~D., {Wells} M., {et al.}, 2003, \aj, 125, 2361

\bibitem[{{Berta} {et~al.}(2007){Berta}, {Lonsdale}, {Siana}, \& {et
  al.}}]{berta07}
{Berta} S., {Lonsdale} C.~J., {Siana} B., {et al.}, 2007, \aap, 467, 565

\bibitem[{{Bertin} \& {Arnouts}(1996)}]{bertin1996117393}
{Bertin} E., {Arnouts} S., 1996, \aaps, 117, 393

\bibitem[{{Blain} {et~al.}(2003){Blain}, {Barnard}, \& {Chapman}}]{blain03}
{Blain} A.~W., {Barnard} V.~E., {Chapman} S.~C., 2003, \mnras, 338, 733

\bibitem[{{Caputi} {et~al.}(2007){Caputi}, {Lagache}, {Yan}, \& {et
  al.}}]{caputi07}
{Caputi} K.~I., {Lagache} G., {Yan} L., {et al.}, 2007, \apj, 660, 97

\bibitem[{{Chapman} {et~al.}(2005){Chapman}, {Blain}, {Smail}, \&
  {Ivison}}]{chapman05}
{Chapman} S.~C., {Blain} A.~W., {Smail} I., {Ivison} R.~J., 2005, \apj, 622,
  772

\bibitem[{{Chary} \& {Elbaz}(2001)}]{chary01}
{Chary} R., {Elbaz} D., 2001, \apj, 556, 562

\bibitem[{{Clements} {et~al.}(2010{\natexlab{a}}){Clements}, {Bendo},
  {Pearson}, {Khan}, {Matsuura}, \& {Shirahata}}]{clements10b}
{Clements} D.~L., {Bendo} G., {Pearson} C., {Khan} S.~A., {Matsuura} S.,
  {Shirahata} M., 2010{\natexlab{a}}, \mnras, 1778

\bibitem[{{Clements} {et~al.}(2010{\natexlab{b}}){Clements}, {Dunne}, \&
  {Eales}}]{clements10}
{Clements} D.~L., {Dunne} L., {Eales} S., 2010{\natexlab{b}}, \mnras, 403, 274

\bibitem[{{Clements} {et~al.}(2008){Clements}, {Vaccari}, {Babbedge}, \& {et
  al}}]{clements08}
{Clements} D.~L., {Vaccari} M., {Babbedge} T., {et al}, 2008, \mnras, 387, 247

\bibitem[{{Coppin} {et~al.}(2008){Coppin}, {Halpern}, {Scott}, \& {et
  al.}}]{coppin08}
{Coppin} K., {Halpern} M., {Scott} D., {et al.}, 2008, \mnras, 384, 1597

\bibitem[{{Donley} {et~al.}(2008){Donley}, {Rieke}, {P{\'e}rez-Gonz{\'a}lez},
  \& {Barro}}]{donley08}
{Donley} J.~L., {Rieke} G.~H., {P{\'e}rez-Gonz{\'a}lez} P.~G., {Barro} G.,
  2008, \apj, 687, 111

\bibitem[{{Draine}(2003)}]{draine03}
{Draine} B.~T., 2003, \araa, 41, 241

\bibitem[{{Dunne} {et~al.}(2000){Dunne}, {Eales}, {Edmunds}, {Ivison},
  {Alexander}, \& {Clements}}]{dunne00}
{Dunne} L., {Eales} S., {Edmunds} M., {Ivison} R., {Alexander} P., {Clements}
  D.~L., 2000, \mnras, 315, 115

\bibitem[{{Dunne} \& {Eales}(2001)}]{dunne01}
{Dunne} L., {Eales} S.~A., 2001, \mnras, 327, 697

\bibitem[{{Dye} {et~al.}(2008){Dye}, {Eales}, {Aretxaga}, \& {et al.}}]{dye08}
{Dye} S., {Eales} S.~A., {Aretxaga} I., {et al.}, 2008, \mnras, 386, 1107

\bibitem[{{Farrah} {et~al.}(2003){Farrah}, {Afonso}, {Efstathiou},
  {Rowan-Robinson}, {Fox}, \& {Clements}}]{farrah03}
{Farrah} D., {Afonso} J., {Efstathiou} A., {Rowan-Robinson} M., {Fox} M.,
  {Clements} D., 2003, \mnras, 343, 585

\bibitem[{{Fixsen} {et~al.}(1998){Fixsen}, {Dwek}, {Mather}, {Bennett}, \&
  {Shafer}}]{fixsen98}
{Fixsen} D.~J., {Dwek} E., {Mather} J.~C., {Bennett} C.~L., {Shafer} R.~A.,
  1998, \apj, 508, 123

\bibitem[{{Franceschini} {et~al.}(2001){Franceschini}, {Aussel}, {Cesarsky},
  {Elbaz}, \& {Fadda}}]{franceschini01}
{Franceschini} A., {Aussel} H., {Cesarsky} C.~J., {Elbaz} D., {Fadda} D., 2001,
  \aap, 378, 1

\bibitem[{{Frayer} {et~al.}(2006){Frayer}, {Fadda}, {Yan}, \& {et
  al.}}]{frayer06}
{Frayer} D.~T., {Fadda} D., {Yan} L., {et al.}, 2006, \aj, 131, 250

\bibitem[{{Genzel} \& {Cesarsky}(2000)}]{genzel00}
{Genzel} R., {Cesarsky} C.~J., 2000, \araa, 38, 761

\bibitem[{{Gispert} {et~al.}(2000){Gispert}, {Lagache}, \& {Puget}}]{gispert00}
{Gispert} R., {Lagache} G., {Puget} J.~L., 2000, \aap, 360, 1

\bibitem[{{Hauser} \& {Dwek}(2001)}]{hauser01}
{Hauser} M.~G., {Dwek} E., 2001, \araa, 39, 249

\bibitem[{{Hildebrand}(1983)}]{hildebrand83}
{Hildebrand} R.~H., 1983, \qjras, 24, 267

\bibitem[{{Ho} {et~al.}(1997){Ho}, {Filippenko}, \& {Sargent}}]{ho97}
{Ho} L.~C., {Filippenko} A.~V., {Sargent} W.~L.~W., 1997, \apj, 487, 579

\bibitem[{{Kartaltepe} {et~al.}(2010){Kartaltepe}, {Sanders}, {Le Floc'h}, \&
  {et al.}}]{kartaltepe10}
{Kartaltepe} J.~S., {Sanders} D.~B., {Le Floc'h} E., {et al.}, 2010, \apj, 709,
  572

\bibitem[{{Kauffmann} {et~al.}(2003){Kauffmann}, {Heckman}, {Tremonti}, \& {et
  al.}}]{kauffmann03}
{Kauffmann} G., {Heckman} T.~M., {Tremonti} C., {et al.}, 2003, \mnras, 346,
  1055

\bibitem[{{Kewley} {et~al.}(2006){Kewley}, {Groves}, {Kauffmann}, \&
  {Heckman}}]{kewley06}
{Kewley} L.~J., {Groves} B., {Kauffmann} G., {Heckman} T., 2006, \mnras, 372,
  961

\bibitem[{{Kewley} {et~al.}(2001){Kewley}, {Heisler}, {Dopita}, \&
  {Lumsden}}]{kewley01}
{Kewley} L.~J., {Heisler} C.~A., {Dopita} M.~A., {Lumsden} S., 2001, \apjs,
  132, 37

\bibitem[{{Kim} {et~al.}(1998){Kim}, {Veilleux}, \& {Sanders}}]{kim98}
{Kim} D., {Veilleux} S., {Sanders} D.~B., 1998, \apj, 508, 627

\bibitem[{{Klaas} {et~al.}(2001){Klaas}, {Haas}, {M{\"u}ller}, \& {et
  al.}}]{klaas01}
{Klaas} U., {Haas} M., {M{\"u}ller} S.~A.~H., {et al.}, 2001, \aap, 379, 823

\bibitem[{{Kov{\'a}cs} {et~al.}(2006){Kov{\'a}cs}, {Chapman}, {Dowell}, \& {et
  al.}}]{kovacs06}
{Kov{\'a}cs} A., {Chapman} S.~C., {Dowell} C.~D., {et al.}, 2006, \apj, 650,
  592

\bibitem[{{Lacy} {et~al.}(2004){Lacy}, {Storrie-Lombardi}, {Sajina}, \& {et
  al.}}]{lacy04}
{Lacy} M., {Storrie-Lombardi} L.~J., {Sajina} A., {et al.}, 2004, \apjs, 154,
  166

\bibitem[{{Le Floc'h} {et~al.}(2005){Le Floc'h}, {Papovich}, {Dole}, \& {et
  al.}}]{lefloch05}
{Le Floc'h} E., {Papovich} C., {Dole} H., {et al.}, 2005, \apj, 632, 169

\bibitem[{{Lonsdale} {et~al.}(2004){Lonsdale}, {Polletta}, {Surace}, \& {et
  al.}}]{lonsdale04}
{Lonsdale} C., {Polletta} M.~d.~C., {Surace} J., {et al.}, 2004, \apjs, 154, 54

\bibitem[{{Lonsdale} {et~al.}(2003){Lonsdale}, {Smith}, {Rowan-Robinson}, \&
  {et al.}}]{lonsdale03}
{Lonsdale} C.~J., {Smith} H.~E., {Rowan-Robinson} M., {et al.}, 2003, \pasp,
  115, 897

\bibitem[{{Mullaney} {et~al.}(2010){Mullaney}, {Alexander}, {Huynh},
  {Goulding}, \& {Frayer}}]{mullaney10}
{Mullaney} J.~R., {Alexander} D.~M., {Huynh} M., {Goulding} A.~D., {Frayer} D.,
  2010, \mnras, 401, 995

\bibitem[{{P{\'e}rez-Gonz{\'a}lez} {et~al.}(2005){P{\'e}rez-Gonz{\'a}lez},
  {Rieke}, {Egami}, \& {et al.}}]{perezgonzalez05}
{P{\'e}rez-Gonz{\'a}lez} P.~G., {Rieke} G.~H., {Egami} E., {et al.}, 2005,
  \apj, 630, 82

\bibitem[{{Pierre} {et~al.}(2007){Pierre}, {Chiappetti}, {Pacaud}, \& {et
  al.}}]{pierre07}
{Pierre} M., {Chiappetti} L., {Pacaud} F., {et al.}, 2007, \mnras, 382, 279

\bibitem[{{Pilbratt} {et~al.}(2010){Pilbratt}, {Riedinger}, {Passvogel}, \& {et
  al.}}]{pilbratt10}
{Pilbratt} G.~L., {Riedinger} J.~R., {Passvogel} T., {et al.}, 2010, \aap, 518,
  L1+

\bibitem[{{Planck Collaboration}(2011)}]{planckcolabclements11}
{Planck Collaboration}, 2011, {ArXiv e-prints, arXiv:1101.2045}

\bibitem[{{Puget} {et~al.}(1996){Puget}, {Abergel}, {Bernard}, {Boulanger},
  {Burton}, {Desert}, \& {Hartmann}}]{puget96}
{Puget} J., {Abergel} A., {Bernard} J., {Boulanger} F., {Burton} W.~B.,
  {Desert} F., {Hartmann} D., 1996, \aap, 308, L5+

\bibitem[{{Rowan-Robinson} {et~al.}(2008){Rowan-Robinson}, {Babbedge},
  {Oliver}, \& {et al.}}]{mrr08}
{Rowan-Robinson} M., {Babbedge} T., {Oliver} S., {et al.}, 2008, \mnras, 386,
  697

\bibitem[{{Rowan-Robinson} {et~al.}(2005){Rowan-Robinson}, {Babbedge},
  {Surace}, \& {et al.}}]{mrr05}
{Rowan-Robinson} M., {Babbedge} T., {Surace} J., {et al.}, 2005, \aj, 129, 1183

\bibitem[{{Rowan-Robinson} \& {Crawford}(1989)}]{mrr89}
{Rowan-Robinson} M., {Crawford} J., 1989, \mnras, 238, 523

\bibitem[{{Rowan-Robinson} {et~al.}(2004){Rowan-Robinson}, {Lari},
  {Perez-Fournon}, \& {et al.}}]{mrr04}
{Rowan-Robinson} M., {Lari} C., {Perez-Fournon} I., {et al.}, 2004, \mnras,
  351, 1290

\bibitem[{{Rowan-Robinson} {et~al.}(2010){Rowan-Robinson}, {Roseboom},
  {Vaccari}, \& {et al.}}]{mrr10}
{Rowan-Robinson} M., {Roseboom} I.~G., {Vaccari} M., {et al.}, 2010, \mnras,
  409, 2

\bibitem[{{Sanders} \& {Mirabel}(1996)}]{sanders96}
{Sanders} D.~B., {Mirabel} I.~F., 1996, \araa, 34, 749

\bibitem[{{Scott} {et~al.}(2002){Scott}, {Fox}, {Dunlop}, \& {et
  al.}}]{scott02}
{Scott} S.~E., {Fox} M.~J., {Dunlop} J.~S., {et al.}, 2002, \mnras, 331, 817

\bibitem[{{Smail} {et~al.}(1997){Smail}, {Ivison}, \& {Blain}}]{smail97}
{Smail} I., {Ivison} R.~J., {Blain} A.~W., 1997, \apjl, 490, L5+

\bibitem[{{Soifer} {et~al.}(2008){Soifer}, {Helou}, \& {Werner}}]{soifer08}
{Soifer} B.~T., {Helou} G., {Werner} M., 2008, \araa, 46, 201

\bibitem[{{Surace} {et~al.}(2005){Surace}, {Shupe}, {Fang}, \& {SWIRE
  Team}}]{surace05}
{Surace} J.~A., {Shupe} D.~L., {Fang} F., {SWIRE Team}, 2005, AAS, 37, 1246

\bibitem[{{Symeonidis} {et~al.}(2009){Symeonidis}, {Page}, {Seymour}, \& {et
  al.}}]{symeonidis09}
{Symeonidis} M., {Page} M.~J., {Seymour} N., {et al.}, 2009, \mnras, 397, 1728
  (S09)

\bibitem[{{Symeonidis} {et~al.}(2007){Symeonidis}, {Rigopoulou}, {Huang}, \&
  {et al.}}]{symeonidis07}
{Symeonidis} M., {Rigopoulou} D., {Huang} J., {et al.}, 2007, \apjl, 660, L73

\bibitem[{{Symeonidis} {et~al.}(2010){Symeonidis}, {Rosario}, {Georgakakis}, \&
  {et al.}}]{symeonidis10}
{Symeonidis} M., {Rosario} D., {Georgakakis} A., {et al.}, 2010, \mnras, 403,
  1474

\bibitem[{{Symeonidis} {et~al.}(2008){Symeonidis}, {Willner}, {Rigopoulou}, \&
  {et al.}}]{symeonidis08}
{Symeonidis} M., {Willner} S.~P., {Rigopoulou} D., {et al.}, 2008, \mnras, 385,
  1015

\bibitem[{{Trichas} {et~al.}(2009){Trichas}, {Georgakakis}, {Rowan-Robinson},
  {Nandra}, {Clements}, \& {Vaccari}}]{trichas09}
{Trichas} M., {Georgakakis} A., {Rowan-Robinson} M., {Nandra} K., {Clements}
  D., {Vaccari} M., 2009, \mnras, 399, 663

\bibitem[{{Trichas} {et~al.}(2010){Trichas}, {Rowan-Robinson}, {Georgakakis},
  \& {et al.}}]{trichas10}
{Trichas} M., {Rowan-Robinson} M., {Georgakakis} A., {et al.}, 2010, \mnras,
  405, 2243

\bibitem[{{Veilleux} {et~al.}(1999){Veilleux}, {Kim}, \&
  {Sanders}}]{veilleux99}
{Veilleux} S., {Kim} D., {Sanders} D.~B., 1999, \apj, 522, 113

\bibitem[{{Veilleux} {et~al.}(1995){Veilleux}, {Kim}, {Sanders}, {Mazzarella},
  \& {Soifer}}]{veilleux95}
{Veilleux} S., {Kim} D., {Sanders} D.~B., {Mazzarella} J.~M., {Soifer} B.~T.,
  1995, \apjs, 98, 171

\bibitem[{{Vlahakis} {et~al.}(2005){Vlahakis}, {Dunne}, \&
  {Eales}}]{vlahakis05}
{Vlahakis} C., {Dunne} L., {Eales} S., 2005, \mnras, 364, 1253

\bibitem[{{Werner} {et~al.}(2004){Werner}, {Roellig}, {Low}, \& {et
  al.}}]{werner04}
{Werner} M.~W., {Roellig} T.~L., {Low} F.~J., {et al.}, 2004, \apjs, 154, 1

\bibitem[{{Yang} {et~al.}(2007){Yang}, {Greve}, {Dowell}, \& {Borys}}]{yang07}
{Yang} M., {Greve} T.~R., {Dowell} C.~D., {Borys} C., 2007, \apj, 660, 1198

\end{thebibliography}

\end{document}